\documentclass[aps,prb,twocolumn,amsmath,amssymb,superscriptaddress,scrartcl,eqsecnum,longbibliography,nofootinbib]{revtex4-2}

\usepackage[T1]{fontenc}
\usepackage[latin9]{inputenc}
\usepackage{color}
\usepackage{verbatim}
\usepackage{amsmath}
\usepackage{amssymb}
\usepackage{graphicx}

\usepackage{bm}

\usepackage{tikz}
\usepackage{diagbox}
\usepackage[normalem]{ulem}

\makeatletter
\PassOptionsToPackage{caption=false}{subfig} 
\usepackage{hyperref}
\hypersetup{
breaklinks=true,
colorlinks=true,
citecolor=blue,
linkcolor=black,
filecolor=black,
urlcolor=blue
}
\IfFileExists{lmodern.sty}{\usepackage{lmodern}}{}

\renewcommand{\bar}{\overline}
\renewcommand{\tilde}{\widetilde}

\renewcommand{\Re}{\operatorname{Re}}
\renewcommand{\Im}{\operatorname{Im}}

\newcommand{\SU}{\operatorname{SU}}
\newcommand{\SO}{\operatorname{SO}}
\newcommand{\SL}{\operatorname{SL}}

\makeatletter

\newcommand*{\wideboxed}[1]{\setlength{\fboxsep}{1ex}%
  \fbox{\m@th$\displaystyle#1$}}
\makeatother

\def\be{\begin{equation}}
\def\ee{\end{equation}}

\makeatother

\begin{document}

\title{Phase Transitions in Quasi-Periodically
Driven
Quantum\\ Critical Systems: Analytical Results }

\author{Jiyuan Fang}
\affiliation{School of Physics, Georgia Institute of Technology, Atlanta, GA 30332, USA}

\author{Qi Zhou}

\affiliation{Chern Institute of Mathematics and LPMC, Nankai University, Tianjin, 300071, China}

\author{Xueda Wen}
\affiliation{School of Physics, Georgia Institute of Technology, Atlanta, GA 30332, USA}

\begin{abstract}
In this work, we study \textit{analytically} the phase transitions in quasi-periodically driven one dimensional quantum critical systems that are described by conformal field theories (CFTs). The phase diagrams and phase transitions can be analytically obtained by using Avila's global theory in one-frequency quasiperiodic cocycles.
Compared to the previous works where the quasiperiodicity was introduced in the driving time and no phase transitions were observed \cite{wen2020periodically}, here we propose a setup where the quasiperiodicity is introduced in the driving Hamiltonians.
In our setup, one can observe the heating phases, non-heating phases, and the phase transitions. The phase diagram as well as the Lyapunov exponents that determine the entanglement entropy evolution can be analytically obtained. In addition, based on Avila's theory, we prove there is no phase transition in the previously proposed setup of quasi-periodically driven CFTs \cite{wen2020periodically}. We verify our field theory results by studying the time evolution of entanglement entropy on lattice models.

\end{abstract}
\maketitle

\tableofcontents

\section{Introduction}
\label{Sec:Introduction}

Quantum phase transitions play an important role in our understanding of fundamental physical properties such as the universality \cite{Sachdev_2011}. Despite of the different microscopic details, two different systems such as water and magnet can behave essentially in the same way in the large length and time scales near the phase transition. Such universal behavior allows us to understand the properties of complicated materials by studying a simpler toy model. Many progresses have been made in the study of phase transitions when the system is in a thermal equilibrium \cite{Sachdev_2011}. When the system is out of equilibrium, there could still be phase transitions with richer features, but are less well understood comparing to their equilibrium counterparts. To understand the nature of non-equilibrium phase transitions in quantum many-body systems, analytically solvable toy models or setups would be very valuable.

In this work, we are interested in the non-equilibrium quantum phase transitions in one dimensional quantum critical systems under a \textit{quasiperiodic} driving, which is analytically solvable. It has been known that quasi-periodically driven quantum systems provide a rich source of intriguing non-equilibrium physics, including  
the quasi-time crystals \cite{2018_quasi_timecrystal,2019_quasi-time_crystal,2019_Zhao_Mintert_Knolle}, 
dynamical localization generated by quasiperiodic drivings \cite{1984_PRA,1994_exp_dynamical_Localization}, and
quasi-periodically driven topological systems \cite{2017_Martin,2018_Peng,2020_Else,2022_Potter}.
 In general, the phase diagrams in quasi-periodically driven quantum many-body systems are very rich, but at the same time are difficult to study -- one usually has to rely on numerical studies \cite{wen2020periodically,2020Lapierre,2024_Schmid}.
In this work, we will introduce a setup where the quasiperiodic dynamics is analytically solvable, by making a connection to Avila's global theory in one-frequency quasiperiodic cocycles  \cite{Avila_global}, one of his Fields Medal work.
This allows us to obtain analytically the phase diagram including the heating and non-heating phases, as well as the growing rates of entanglement entropy evolution in the heating phase.

\subsection{Background of time-dependent driven CFTs}

The setup we will study is based on the recently developed time-dependent driven conformal field theories, which are exactly solvable \cite{wen2018floquet,Wen_2018,Fan_2020,Lapierre:2019rwj,lapierre2020geometric,fan2020General, han2020classification,2020Lapierre,wen2020periodically, 2021Ageev,2020_realtime,2021Das,RandomCFT2021,
2022Das_OTOC,2022Bermond,2022Choo,2022Cooling,2023_OBC,2023_StatePrepare,2024Ryu,2023Das,2023_Nozaki_Scrambling, 2024_Krylov,2024_Guo}.
This setup can be realized by a deformation of Hamiltonian density \textit{both in time and in space}:
\be
\label{H_deform_intro}
H_{\text{deform}}(t)=\int dx \, f(x,t)\, T_{00}(x),
\ee
where $T_{00}(x)$ is the Hamiltonian density before the deformation, and $f(x,t)$ is a smooth and real function in space $x$, but it is not necessarily smooth in time $t$, which will be clear in our setup later. The choice of $f(x)$ determines the underlying algebra in the time-dependent drivings.
For a general choice of $f(x,t)$
in \eqref{H_deform_intro}, the generators for the driving Hamiltonians will form an infinite dimensional Virasoro algebra. However, if we choose $f(x,t)$ in the simple form
\be
\label{fx_sl2}
f(x,t)=a^+(t)\cdot\cos\left(\frac{2\pi x}{l}\right)+a^-(t)\cdot \sin\left(\frac{2\pi x}{l}\right)+a^0(t),
\ee
where $a^+,\, a^-,\,a^0\in \mathbb R$ and $l$ is a characteristic length of the deformation,\footnote{The reason we use the notations $a^+$, $a^-$, and $a^0$ will become clear later in \eqref{Hdeform_SL2}.}
then the generators of the driving Hamiltonians will form a finite dimensional $sl(2,\mathbb R)$ algebra, which is a sub-algebra of the Virasoro algebra. See Sec.\ref{Sec:drivenCFT_general} for more details.
In the following discussions, we will refer to \textit{general deformations} as a general choice of $f(x,t)$ where the underlying algebra is the infinitely dimensional Virasoro algebra, and refer to \textit{$sl(2,\mathbb R)$ deformations} as the specific choice in \eqref{fx_sl2}, where the underlying algebra is the finite $sl(2,\mathbb R)$ algebra.

\medskip
Till now, there have been extensive studies on the properties of time-dependent driven CFTs by using the deformed Hamiltonians of the form in \eqref{H_deform_intro}. In general, two different non-equilibrium phases can be observed: \textit{heating phase} and \textit{non-heating phase}, with a phase transition in between.
The ``order parameter'' used to distinguish these two phases are the time evolutions of entanglement entropy or total energy.
In the heating phase, the total energy of the system keeps growing exponentially in time, with the absorbed energy mainly accumulated at certain ``hot spots'' in real space\cite{Fan_2020,Lapierre:2019rwj}. The entanglement entropy
of a subsystem that contains such hot spots grows linearly in time\cite{wen2018floquet}. In particular, the entanglement is mainly contributed by
the neighboring hot spots which are strongly entangled with each other \cite{Fan_2020,wen2020periodically,RandomCFT2021}.
In the non-heating phase, both the entanglement entropy and energy density will oscillate in time. At the phase transition, one can find the entanglement entropy grows logarithmically in time, while the total energy grows in a power law in time\cite{wen2018floquet,Fan_2020,Lapierre:2019rwj,wen2020periodically}.

\medskip

In the following, let us give a brief review of recent understanding of the phase diagrams in periodically, quasi-periodically, and randomly driven CFTs, respectively.

\medskip
-- \textit{Periodically driven CFTs}:

The periodically CFTs (Floquet CFTs), have been studied with both $sl(2,\mathbb R)$ deformations
\cite{wen2018floquet,Fan_2020,Lapierre:2019rwj,han2020classification,2020Lapierre,wen2020periodically,2020_realtime,2021Das}
and general deformations\cite{lapierre2020geometric,fan2020General}. In both cases, one can in general observe the two phases introduced above in the phase diagram. These phases are robust even if the initial state is chosen as a thermal ensemble\cite{2022Cooling,2022Choo}, since it is the operator evolution, which is independent of the initial state, that determines the phase diagram.
More explicitly, in a Floquet CFT with $sl(2,\mathbb R)$ deformations, the phase is determined by the types of operator evolution within a single driving cycle. The heating, non-heating phases, and phase transitions correspond to the three different types of M\"obius transformations, i.e., hyperbolic, elliptic, and parabolic, respectively\cite{wen2018floquet}.
In a Floquet CFT with general deformations, the phase is determined by the trajectory of operator evolution in time:\cite{lapierre2020geometric,fan2020General} 
the presence of stable fixed points in the operator evolution indicates that the driven CFT is in a heating phase. If the fixed points become critical, which typically arise due to the merging of a pair of stable and unstable fixed points, then the driven CFT is at the phase transition. Otherwise, if there are no fixed points at all, then the driven CFT is in a non-heating phase.

\begin{table}
\centering
\small
\def\arraystretch{1.5}
\begin{tabular}{| c|c|c|c |}
\hline
    Driven CFTs  & Heating    & Phase Transition & Non-heating \\
    \hline
    \shortstack{Periodic\\
\cite{wen2018floquet,Fan_2020,Lapierre:2019rwj,lapierre2020geometric,fan2020General,han2020classification,wen2020periodically}} & $\surd$   & $\surd$  & $\surd$  \\
    \hline
   Random\cite{2022RandomCFT}  & $\surd$   & $\times$  & $\times$ \\
    \hline
\shortstack{ Quasi-periodic \\ Fibonacci\cite{wen2020periodically,2020Lapierre}} & $\surd$   & $\surd$ & $\times$ \\
    \hline
    \shortstack{ Quasi-periodic \\Aubry-Andre\cite{wen2020periodically}} & $\surd$    & $\times$ & $\times$ \\
    \hline
        \shortstack{ \textcolor{black}{Quasi-periodic} \\ \textcolor{black}{[This work]}} & \textcolor{black}{$\surd$}    & \textcolor{black}{$\surd$} & \textcolor{black}{$\surd$} \\
    \hline
\end{tabular}
\caption{Non-equilibrium phases and phase transitions in a time-dependent driven CFT with $sl(2,\mathbb R)$ deformations for different types of drivings. Here the symbol ``$\surd$'' (``$\times$'') indicates that the corresponding non-equilibrium phase exists (does not exist).}
\label{Phase_table}
\end{table}

\medskip
-- \textit{Quasi-periodically driven CFTs}:

The quasi-periodically driven CFTs were studied only with the $sl(2,\mathbb R)$ deformations\cite{wen2020periodically,2020Lapierre}.
Till now, two types of quasi-periodical drivings in the driven CFTs  were studied: the Fibonacci driving\cite{wen2020periodically,2020Lapierre}, and the Aubry-Andre driving\cite{wen2020periodically}. Interestingly, only the heating phase and the critical-point feature  were observed in these two setups.
More explicitly, in the Fibonacci driving, where the driving sequence is generated by a Fibonacci bitstring, it was found that the system is in a heating phase for most choices of driving parameters. At certain parameters, one can observe the critical-point feature:
the entanglement entropy/energy oscillate at the Fibonacci numbers, but grow logarithmically/polynomially at the non-Fibonacci numbers. In the other setup on quasi-periodic driving, where one considered an Aubry-Andre like sequence\cite{wen2020periodically}, which we will also study in this work, only the heating phase was observed based on a numerical study.

\medskip
-- \textit{Randomly driven CFTs}:

The randomly driven CFTs were also only studied with the $sl(2,\mathbb R)$ deformations. As shown in Ref.\cite{RandomCFT2021}, the fate of a randomly driven CFT with $sl(2,\mathbb R)$ deformations can be understood based on Furstenberg's theorem on the random products of $\SL(2,\mathbb R)$ matrices\cite{Furstenberg1963}. One can prove that for most choices of parameters, which satisfy Furstenberg's criteria, the driven CFT is in a heating phase. For those driving parameters that do not satisfy Furstenberg's criteria, it is found the driven CFT is at the ``exceptional point'', where the entanglement entropy grows in time as $\sqrt t$, but the total energy still grows exponentially in time. In short, a randomly driven CFT with $sl(2,\mathbb R)$ deformations has been completely classified and characterized in \cite{RandomCFT2021}. 

A summary of the phase diagrams as briefly reviewed above is given in Table \ref{Phase_table}.

\medskip

Besides the phase diagrams and their characterizations in driven CFTs, we want to emphasize that there were various developments and generalizations in this direction \cite{2021Nozaki,2023Caputa,2023_Geometry, 2023_Briding,2024_Ge,2024Nozaki2,2024Mezei,2023_Brane,2024_Modular,2024_Bai_Miyata_Nozaki,
2024_Das,
HigherFloquetCFT,2024Krylov,2024Lapierre,2024Nozaki,2024JSM_Wen,2023Liu_quench,2408_Moosavi,2024_Asymmetry_Floquet}, some of which we briefly review below. 

In \cite{2021Nozaki,2023Caputa,2023_Geometry, 2023_Briding,2024_Ge,2024Nozaki2,2024Mezei,2023_Brane,2024_Modular,2024_Bai_Miyata_Nozaki,2024_Das}, the holographic dual of inhomogeneous quantum quenches as well as Floquet CFTs with $sl(2,\mathbb R)$ deformations were studied. For example, in \cite{2023_Geometry}, it was found that the time-dependent driven CFTs are dual to driven BTZ black holes with time-dependent horizons. In the heating phase of Floquet CFTs, the black hole horizon approaches the boundary at certain specific points, which correspond to the hot spots in the driven CFT \cite{Fan_2020,Lapierre:2019rwj,wen2020periodically,2022Cooling}, where the entropy and the absorbed energy are accumulated during the driving. In particular, the horizon approaches such hot spots at a linear rate,  which is given by the Lyapunov exponent that measures the entanglement entropy growth in the driven CFT.
In the non-heating phase, the black hole horizon simply oscillates in time. Similar phenomena also appear in the holographic dual of inhomogeneous quantum quenches with $sl(2,\mathbb R)$ deformations. See, e.g., \cite{2021Nozaki} for more details.

The time-dependent driven CFTs also provide a nice platform to illustrate the quantum complexity in quantum field theories \cite{2022_Caputa,2023_Geometry,2024Krylov}. Briefly, the notion of complexity as defined in quantum information theory characterizes the minimal number of unitary gates that are required to reach a specific state from a reference state.
In \cite{2024Krylov}, the so-called Krylov complexity, which characterizes the growth of a local operator in operator Hilbert space, was studied in a Floquet CFT. It was found that the Krylov complexity grows exponentially in time in the heating phase, oscillates in the non-heating phase, and grows polynomially at the phase transition. A different definition of complexity and its application in Floquet CFT was studied earlier in \cite{2023_Geometry}, and the complexity displays different features in different phases -- it grows linearly in time in the heating phase and oscillates in time in the non-heating phase.

There are also various interesting applications of the driven CFTs. In \cite{2022Cooling}, it was found that for a given Gibbs state at finite temperature, one can perform a conformal cooling for a target subsystem by tuning driving parameters to the heating phase of a Floquet CFT. The prescribed subsystem can be cooled down to zero temperature exponentially rapidly in time. 
See also \cite{2021Nozaki} for the conformal cooling in CFTs by using an inhomogeneous quantum quench.
In \cite{2024Lapierre}, it was found that the Floquet driving can be used to engineer the inhomogeneous quantum chaos in CFTs. By tuning the driving parameters, one can realize a transition from chaotic to non-chaotic regimes in the quantum critical systems.

There are also some other very interesting progresses in the time-dependent driven CFTs and inhomogeneous quenches by deforming the Hamiltonian density, which we will not review here. These include the non-unitary time evolution in inhomogeneous quantum quenches and Floquet CFTs \cite{2024Nozaki,2024JSM_Wen}, 
the effect of boundary conditions in the inhomogeneous quantum quenches with $sl(2,\mathbb R)$ deformations \cite{2023Liu_quench}, the perfect wave transfer in inhomogeneous CFTs \cite{2408_Moosavi}, and features of entanglement asymmetry in a Floquet CFT \cite{2024_Asymmetry_Floquet}.

The works reviewed so far are mainly in (1+1) dimensions. 
Very recently in \cite{HigherFloquetCFT}, the authors generalized the Floquet CFTs to $(d+1)$ dimensions with arbitrary $d>1$, where the Floquet dynamics can be exactly solvable. 
By deforming the Hamiltonian density in space, the driving Hamiltonians are linear combinations of generators for $\SL(2,\mathbb R)$, which is a subgroup of the conformal group $\SO(d+2,1)$ in $(d+1)$ dimensional CFTs. With this construction, one can obtain the heating phases, non-heating phases, and the phase transitions during the Floquet driving.

Last but not least, we hope to emphasize the early works on inhomogeneous quantum quenches \cite{2016_Dubail,2017_Dubail,2017_luttinger,2018_GLM,2019_Moosavi}, although their motivations are not on the non-equilibrium phase transitions. In these works, by considering a certain deformation of the Hamiltonian density in space, the quench dynamics may be analytically solvable by mapping the problem to CFTs in curved space. While in driven CFTs, as we have reviewed above, we are interested in the rich phase diagrams where different non-equilibrium phases can emerge during the time-dependent driving.

\subsection{Motivations and main results}

\begin{figure}
\centering
	\begin{tikzpicture}[baseline={(current bounding box.center)}]
    \normalsize
	\node at (16pt, 45pt){\underline{Type-I driving:}};
	\small    
	\node at (-5pt, 7pt){$H(t)$:};
	\draw [thick](20pt,0pt)--(40pt,0pt) --(40pt,20pt) -- (60pt,20pt);
	
	\draw [thick](60pt,20pt)--(60pt,0pt) --(100pt,0pt) --(100pt,20pt) -- (120pt,20pt);
	
	\draw [thick](120pt,20pt)--(120pt,0pt)--(180pt,0pt) -- (180pt,20pt) -- (200pt,20pt);
	
	\node at (10+40pt,28pt){$H_1$};
	\node at (30pt, 8pt){$H_0$};

 \node at (30+50pt, 8pt){$H_0$};
 	\node at (10+100pt,28pt){$H_1$};

  \node at (30+120pt, 8pt){$H_0$}; 
 	\node at (10+180pt,28pt){$H_1$};

	\draw[>=stealth,<->] (20pt,-5pt) --node[below]{$\omega l$} (40pt,-5pt);
	\draw[>=stealth,<->] (60pt,-5pt) --node[below]{$2\omega l$} (100pt,-5pt);
	\draw[>=stealth,<->] (120pt,-5pt) --node[below]{$3\omega l$} (180pt,-5pt);
	
	\draw [>=stealth,->] (90pt, -20pt)--(130pt,-20pt);
	\node at (110pt, -25pt){$\text{time}$};

	\node at (10+200pt,12pt){$\cdots$};

\begin{scope}[xshift=0pt,yshift=-90pt]
\normalsize
	\node at (18pt, 45pt){\underline{Type-II driving:}};
\small
	\node at (-5pt, 10pt){$H(t)$:};
	\draw [thick](20pt,0pt)--(40pt,0pt) --(40pt,20pt) -- (60pt,20pt);
	
	\draw [thick](60pt,20pt)--(60pt,5pt) --(80pt,5pt) --(80pt,20pt) -- (100pt,20pt)--(100pt,12pt)--(120pt,12pt)--(120pt,20pt)--(140pt,20pt)--(140pt,9pt)--(160pt,9pt)--(160pt,20pt)--(180pt,20pt)--(180pt,4pt)--(200pt,4pt);

	\node at (10+40pt,28pt){$H_1$};
 	\node at (10+80pt,28pt){$H_1$};
  	\node at (10+120pt,28pt){$H_1$};
   	\node at (10+160pt,28pt){$H_1$};
    
	\node at (30pt, 8pt){$\textcolor{black}{H_{0,1}}$};
 	\node at (30+40pt, 12pt){$\textcolor{black}{H_{0,2}}$};
  	\node at (30+80pt, 19pt){$\textcolor{black}{H_{0,3}}$};
   	\node at (30+120pt, 16pt){$\textcolor{black}{H_{0,4}}$};
    \node at (30+160pt, 11pt){$\textcolor{black}{H_{0,5}}$};

	\node at (10+200pt,15pt){$\cdots$};

	\draw [>=stealth,->] (90pt, -10pt)--(130pt,-10pt);
	\node at (110pt, -15pt){$\text{time}$};
\end{scope} 
	\end{tikzpicture}
 \caption{
 Two different ways of doing quasiperiodic drivings as considered in this work.
 In the type-I driving (top), we fix the two driving Hamiltonians and introduce the quasi-periodicity in the time interval for $H_0$. 
The time intervals for $H_0$ evolution in the $n$-th step is $n\omega l$, where $\omega$ is an irrational number, while the time intervals for $H_1$ evolution are constant.
 This protocol was numerically studied in Ref.\onlinecite{QuasiPeriodic}. 
 In the type-II driving (bottom), we fix the time intervals of driving in every other step.
The driving Hamiltonian $H_1$ in every other step is fixed, while the driving Hamiltonians $H_{0,n}$ depend on $n$ in a quasiperiodic way. See the main text for more details.
}
 \label{Fig:Driving_Protocol}
\end{figure}

In this work, we will focus on a quasi-periodically driven CFT with $sl(2,\mathbb R)$ deformations. We are mainly interested in the following questions:

\begin{enumerate}

\item For the setups of quasi-periodically driven CFTs that are already studied in the literature\cite{wen2020periodically,2020Lapierre}, only the heating phases were observed. Can one have a setup of quasi-periodically driven CFTs where there are both heating phases and non-heating phases with phase transitions?

\item For a general setup of quasi-periodically driven CFT, e.g., the Aubry-Andre-like driving in Ref.\cite{wen2020periodically},
can one find an analytical way to determine the phase diagram? 
Furthermore, in the heating phase, can one obtain analytically the Lyapunov exponents which determine the entanglement entropy evolution?

\end{enumerate}

In this work, we will give affirmative answers to the questions above, as follows:

\begin{enumerate}

\item 
We generalize the setup of quasi-periodically driven CFTs in \cite{wen2020periodically,2020Lapierre} to the case where the driving Hamiltonians themselves depend on parameters in a quasi-periodical way, which is illustrated in the type-II driving in Fig.\ref{Fig:Driving_Protocol}.\footnote{As a remark, while we are interested in the minimal setup of quasiperiodic drivings, 
one can certainly consider the more general setup by combining type-I and type-II drivings above.}
In this new setup, we find the phase transition between heating and non-heating phases is a generic feature.

\item Both types of quasiperiodic drivings in Fig.\ref{Fig:Driving_Protocol} can be studied based on Avila's global theory for quasiperiodic cocycles.
By applying Avilia's global theory, one can evaluate the Lyapunov exponents as well as the acceleration analytically, based on which one can obtain the phase diagrams of quasi-periodically driven CFTs. For the type-I driving in Fig.\ref{Fig:Driving_Protocol}, we can prove that there is only a heating phase, with no phase transitions and non-heating phases.
For the type-II driving, we find there are both heating and non-heating phases with phase transitions. For both type-I and type-II drivings, the Lyapunov exponents obtained from Avilia's theory agree very well with the numerics for a large range of driving parameters.
\end{enumerate}

In the rest of this introduction, let us introduce briefly Avila's global theory and explain why 
it can be applied to a quasi-periodically driven CFT with $sl(2,\mathbb R)$ deformations.

\subsection{Avila's global theory}
\label{Subsec:Avila}

Avila's global theory in one-frequency quasiperiodic cocycles
is an important progress in the spectral theory of self-adjoint quasiperiodic Schr\"odinger operators \cite{Avila_global,Avila_You_Zhou_2017}.
The main goal of Avila's global theory is that, beyond the local problem of understanding the features in different phases of quasiperiodic systems, one should explain how the phase transitions between these phases occur \cite{Avila_global}.
That is, the goal is to understand the global phase diagram and phase transitions in a quasiperiodic system.

\medskip
Let us give a very brief introduction to Avila's global theory. Suppose that $A$ is an analytic function from the circle $S^1$ to the group $\SL(2,\mathbb R)$,
then we can define the one-frequency analytic quasiperiodic cocycles $(\omega, A)$, which can be seen as a linear skew product:
\be
\begin{split}
(\omega, A):\quad S^1\times R^2 &\rightarrow S^1\times R^2,\\
(x,v)&\mapsto (x+\omega,\, A(x)\cdot v),\\
\end{split}
\ee
where $A$ can be viewed as the $x$-dependent $\SL(2,\mathbb R)$ matrix acting projectively on unit vectors $v$.
The celebrated Avila's global theory gives a classification of one-frequency analytic quasiperiodic $\SL(2,\mathbb R)$ cocycles by two dynamical invariants: Lyapunov exponent and acceleration.

First, the Lyapunov exponent of the cocycle $(\omega,\, A)$ is defined as
\be
\label{lambdaL_Anx}
\lambda_L=\lim_{n\to \infty}\frac{1}{n}\int_{S^1} \log|| A_n(x) || dx,
\ee
where 
\be
\label{An_x}
A_n(x)=A\big(x+(n-1)\omega\big)\cdots A(x+\omega)\cdot A(x).
\ee
Then the \textit{complexified} Lyapunov exponents are obtained by 
\be
\label{Complex_Lyapunov_0}
\lambda_L(\epsilon)=\lim_{n\to \infty}\frac{1}{n}\int \log|| A_n(x+i\epsilon) || dx,
\ee
where $A_n(x+i\epsilon)=A(x+i\epsilon+(n-1)\omega)\cdots
A(x+i\epsilon+\omega)\cdot A(x+i\epsilon)$.
Note that the complexified Lyapunov exponents were first proposed by Herman in \cite{Herman1983}. 

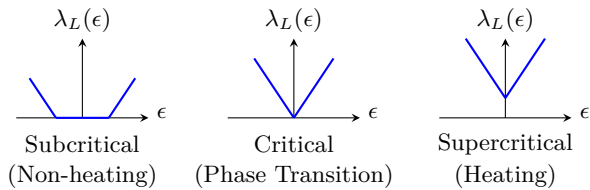
\begin{figure}[t]
\begin{tikzpicture}

\small
    \node at (0pt, -10pt){Subcritical}; 
    \node at (0pt, -22pt){\textcolor{black}{(Non-heating)}}; 
    
         \draw [>=stealth,->] (-25pt, 0pt)--(25pt,0pt);
          \draw [>=stealth,->] (0pt, 0pt)--(0pt,30pt);
                     \node at (30pt, 2pt){$\epsilon$};
                     \node at (0pt, 37pt){$\lambda_L(\epsilon)$};
                     
                              \draw [blue][thick] (-10pt, 0pt)--(10pt,0pt);
                              \draw [blue][thick] (10pt, 0pt)--(20pt,15pt);                     
                                \draw [blue][thick] (-20pt, 15pt)--(-10pt,0pt);    
                                                   
                     \begin{scope}[xshift=80pt]
                         \node at (0pt, -10pt){Critical};   
                             \node at (0pt, -22pt){\textcolor{black}{(Phase Transition)}}; 
    
         \draw [>=stealth,->] (-25pt, 0pt)--(25pt,0pt);
          \draw [>=stealth,->] (0pt, 0pt)--(0pt,30pt);
                     \node at (30pt, 2pt){$\epsilon$};
                     \node at (0pt, 37pt){$\lambda_L(\epsilon)$};
                     
                              \draw [blue][thick] (0pt, 0pt)--(15pt,22.5pt);      
                              \draw [blue][thick] (0pt, 0pt)--(-15pt,22.5pt);                                                  
                     \end{scope}

                     \begin{scope}[xshift=160pt]
                         \node at (0pt, -10pt){Supercritical};           \node at (0pt, -22pt){\textcolor{black}{(Heating)}}; 
      
         \draw [>=stealth,->] (-25pt, 0pt)--(25pt,0pt);
          \draw [>=stealth,->] (0pt, 0pt)--(0pt,30pt);
                     \node at (30pt, 2pt){$\epsilon$};
                     \node at (0pt, 37pt){$\lambda_L(\epsilon)$};
                     
                             \draw [blue][thick] (0pt, 7.5pt)--(15pt,30pt);      
                              \draw [blue][thick] (0pt, 7.5pt)--(-15pt,30pt);    
                     \end{scope}
\end{tikzpicture}
	\caption{
    Complexified Lyapunov exponents $\lambda_L(\epsilon)$ in the subcritical, critical, and supercritical regimes.
    Near $\epsilon=0$, one has $\lambda_L=0$ and $\omega_\lambda=0$ in the subcritical case, $\lambda_L=0$ and $\omega_\lambda\in \mathbb Z^+$ in the critical case, and $\lambda_L>0$ and  $\omega_\lambda\in \mathbb Z^+$ in the supercritical case. In driven CFTs,
    these three cases correspond to the non-heating phase, phase transition, and the heating phase, respectively.
Note the function $\epsilon\mapsto \lambda_L(\epsilon)$ is an even function when the matrices are $\SL(2,\mathbb R)$ or $\SU(1,1)$.
	}
 \label{Lyapunov_Phase}
\end{figure}

Next, the central concept in Avila's global theory is the so-called \textit{acceleration}, which corresponds to the slope of $\lambda_L(\epsilon)$ as \cite{Avila_global}
\be
\label{Acceleration}
\omega_\lambda=\lim_{\epsilon\to 0^+} \frac{\lambda_L(\epsilon)-\lambda_L}{2\pi \epsilon}.
\ee
The key observation in Avila's global theory is that $\epsilon\to \lambda_L(\epsilon)$ is convex and piecewise linear, with the acceleration satisfying 
\be
\omega_\lambda\in \mathbb Z.
\ee
That is, the acceleration $\omega_\lambda$ in \eqref{Acceleration} with an irrational frequency is always quantized to be an integer \cite{Avila_global}.
\footnote{Note that that quantization of $\omega_\lambda$ does not extend to rational frequencies \cite{Avila_global}.}

As we will introduce later in Sec.\ref{Sec:drivenCFT_general}, 
for $sl(2,\mathbb R)$ deformed CFTs, the operator evolutions are determined by the M\"obius transformations that are described by $\SU(1,1)$ matrices. The quasi-periodical driving is realized by first introducing an irrational frequency $\omega$ in such $\SU(1,1)$ matrices and then considering a product of such matrices. Noting that $\SU(1,1)\cong \SL(2,\mathbb R)$, we can apply Avila's theory to the quasi-periodically driven CFTs under $sl(2,\mathbb R)$ deformations.

\medskip
Avila's global theory categorizes the sequence of quasiperiodic cocycles that are \textit{not} uniformly hyperbolic into three cases based on the Lyapunov exponents $\lambda_L$ in \eqref{lambdaL_Anx} and the acceleration $\omega_\lambda$ in \eqref{Acceleration}:

\begin{enumerate}
\item The subcritical regime: $\lambda_L=0$ and $\omega_\lambda=0$.

\item The critical regime: $\lambda_L=0$ and $\omega_\lambda\in \mathbb Z^+$.

\item The supercritical regime: $\lambda_L>0$ and $\omega_\lambda\in \mathbb Z^+$.

\end{enumerate}
Here $\mathbb Z^+$ denote positive integers. Pictorially, the features of the complexified Lyapunov exponents $\lambda_L(\epsilon)$ for these three different cases are shown in Fig.\ref{Lyapunov_Phase}.
One can find that although $\lambda_L=0$ in both
subcritical and critical regimes, it is stable  in the subcritical regime and unstable in the critical regime
under complexification.

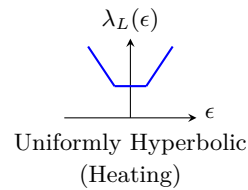
\begin{figure}
\begin{tikzpicture}

\small
         \draw [>=stealth,->] (-25pt, 0pt)--(25pt,0pt);
          \draw [>=stealth,->] (0pt, 0pt)--(0pt,30pt);
                     \node at (30pt, 2pt){$\epsilon$};
                     \node at (0pt, 37pt){$\lambda_L(\epsilon)$};
                     
                              \draw [blue][thick] (-6pt, 12pt)--(6pt,12pt);
                              \draw [blue][thick] (6pt, 12pt)--(16pt,27pt);                     
                                \draw [blue][thick] (-16pt, 27pt)--(-6pt,12pt);    

                         \node at (0pt, -10pt){Uniformly Hyperbolic}; 
                             \node at (0pt, -22pt){\textcolor{black}{(Heating)}};

\end{tikzpicture}
	\caption{Complexified Lyapunov exponents $\lambda_L(\epsilon)$ in the uniformly hyperbolic case, where one has $\lambda_L>0$ and $\omega_\lambda=0$ near $\epsilon=0$. This case corresponds to the heating phase in our driven CFT.
	}
 \label{Fig:Uniform_Hyperbolic}
\end{figure}

As will be discussed in the next sections, in the context of quasi-periodically driven CFTs, these three cases correspond to the non-heating phase, phase transition, and the heating phase, respectively. 

On the other hand, if the quasiperiodic cocycles are uniformly hyperbolic, then one has $\lambda_L>0$ and $\omega_\lambda=0$ \cite{Avila_global}, with the complexified Lyapunov exponents $\lambda_L(\epsilon)$ shown in Fig.\ref{Fig:Uniform_Hyperbolic}.
This case also corresponds to the heating phase in a quasi-periodically driven CFT, since the heating phase therein is only characterized by a positive Lyapunov exponent $\lambda_L>0$.
To distinguish the uniformly and non-uniformly hyperbolic cocycles in the heating phase, one can either check the feature of acceleration $\omega_\lambda$, or check whether 
the exponential growth of $||A_n(x) ||$ is uniform with respect to $x\in S^1$ or not \cite{yoccoz} (See the discussion in Appendix \ref{Uniform_non-uniform}).

Besides categorizing different types of quasiperiodic cocycles, Avila's global theory also provides a very useful tool to 
exactly calculate the Lyapunov exponents. This will be illustrated in Sec.\ref{Sec:NoPhaseTransition} and Sec.\ref{Sec:PhaseTransition}.

\medskip

For readers from condensed matter physics, it is helpful to illustrate the above concepts based on a well known example -- 
the Aubry-Andr\'e model, which is a lattice model with quasiperiodic onsite potential. The Hamiltonian is
\be
H_{\text{AA}}=\sum_n (c_n^\dag c_{n+1}+h.c.)+2\lambda \sum_n \cos(2\pi \omega n+\phi) \, c_n^\dag c_n,
\ee
where $n\in \mathbb Z$ denotes the lattice sites,
the fermionic operators satisfy 
$\{c_i,c_j^\dag\}=\delta_{ij}$, $\{c_i, c_j\}=\{c_i^\dag, c_j^\dag\}=0$,
and $\omega$ is an irrational number.
Here $\lambda>0$ characterizes the strength of the quasiperiodic onsite potentials.
This model has a phase transition at $\lambda=1$ and the two phases (metallic phase for $\lambda<1$ and insulating phase for $\lambda>1$) are related to each other through a duality transformation, which was rigorously proved in mathematics \cite{2014_Damanik,prl1982,SIMON_1982}. 
One can apply Avila's global theory to the $\SL(2,\mathbb R)$ transfer matrix in this model. For all the energy $E$ in the energy spectrum of $H_{\text{AA}}$, the Lyapunov exponents as well as the accelerations defined in \eqref{Acceleration} have the following features: \cite{Avila_global,2002_BJ}
\begin{enumerate}
\item $\lambda<1$: $\lambda_L(E)=0$, and $\omega_\lambda(E)=0$.

\item $\lambda=1$: $\lambda_L(E)=0$, and $\omega_\lambda(E)=1$.

\item $\lambda>1$: $\lambda_L(E)=\log|\lambda|$, and $\omega_\lambda(E)=1$.

\end{enumerate}

These three cases correspond to the subcritical, critical, and supercritical regimes in Fig.\ref{Lyapunov_Phase}, and physically they correspond to the metallic phase, phase transition, and the insulating phase, respectively.


\medskip

The rest of this work is organized as follows.
In Sec.\ref{Sec:drivenCFT_general}, we introduce the general setup of time-dependent driven CFTs with $sl(2,\mathbb R)$ deformations as well as the basic concepts to characterize different non-equilibrium phases. In Sec.\ref{Sec:NoPhaseTransition}, we consider the setup in \cite{wen2020periodically} on quasiperiodic driving and show why there is only a heating phase without any phase transitions. Then in Sec.\ref{Sec:PhaseTransition}, we introduce a new setup of quasi-periodically driven CFTs and study the rich phase diagrams and the features in each phase. 
Then we discuss some future problems and conclude in Sec.\ref{Sec:Discuss}. There are also several appendices. In appendix \ref{Appendix:BuildingBlock}, we give details on the building block of quasiperiodic drivings, i.e., a single quantum quench with different types of quenched Hamiltonians. Then we discuss general cases in type-I and type-II quasiperiodic drivings in Appendix \ref{Appendix:HeatingPhaseGerneral} and Appendix \ref{Appendix:PhaseTransition} respectively.
In appendix \ref{Uniform_non-uniform}, we discuss the difference between uniformly and non-uniformly hyperbolic drivings that will give different features in the sub-leading terms of the entanglement entropy evolutions.

\section{Time-dependent driven CFTs: general setup}
\label{Sec:drivenCFT_general}

In this section, we introduce the general setup of time-dependent driven CFTs, as well as the basic concepts that will be used in the characterization of different phases in the driven CFTs. For more technical details, one can refer to Refs.\cite{wen2020periodically,han2020classification}.

We will mainly focus on the quasiperiodic driving with different types of $sl(2,\mathbb R)$ deformed Hamiltonians as introduced in \eqref{H_deform_intro} and \eqref{fx_sl2}.
It is emphasized that in general one can deform the chiral and anti-chiral Hamiltonian density \textit{independently} as follows:
\footnote{In the numerical calculation on a lattice model, since it is not straightforward to deform the chiral and anti-chiral components of the stress tensor independently, we will always consider the deformation in \eqref{H_deform_intro} and \eqref{fx_sl2}. 
That is, we deform the chiral and anti-chiral components simultaneously.
}
\be\label{Hcft}
H_{\text{CFT}}=H_{\text{chiral}}+H_{\text{anti-chiral}},
\ee
where
\be
\label{Hcfti}
\left\{
\begin{split}
H_{\text{chiral}}=&\frac{1}{2\pi}\int_{0}^{L} f(x)\, T (x) dx,\\
H_{\text{anti-chiral}}=&\frac{1}{2\pi}\int_{0}^{L} g(x)\, \bar T (x) dx.\\
\end{split}
\right.
\ee
Here $T(x)$ ($\bar T(x)$) is the holomorphic (anti-holomorphic), or chiral (anti-chiral) component of the stress tensor.  They are related to the energy density $T_{00}(x)$ and momentum density $T_{01}(x)$ by $T_{00}=(T+\overline{T})/2\pi$ and
$T_{01}=(T-\overline{T})/2\pi$.
Here $f(x)$ and $g(x)$ are the deformation (real) functions that may be independent from each other. By choosing $g(x)=f(x)$, one has the specific deformation in \eqref{H_deform_intro}.
Now let us focus on the effect of deformation in the chiral Hamiltonian $H_{\text{chiral}}$, and the effect of anti-chiral deformation can be similarly discussed.

Considering the deformation function $f(x)$ in the form of 
\eqref{fx_sl2}, then the chiral Hamiltonian in \eqref{Hcfti} can be written in terms of the Virasoro generators as
\be\label{Hdeform_SL2}
H_{\text{chiral}}=\frac{2\pi}{L}\left(a^0 L_0+a^+ L_{q,+}
+a^- L_{q,-}
\right)-\frac{\pi c}{12 L}, 
\ee
where $q=\frac{L}{l}\in\mathbb Z$ characterizes the number of wavelength $l$ in the deformation in \eqref{fx_sl2}, and here we have defined
$L_{q,+}:=(L_q+L_{-q})/2$ and $L_{q,-}:=(L_q-L_{-q})/2i$,
with 
\be
L_{q}:=\frac{c}{24}\delta_{q,0}+\frac{L}{2\pi}\int_0^L \frac{dx}{2\pi}\, e^{i\frac{2\pi q}{L}x} \, T(x)
\ee
being the generators of Virasoro algebra 
\be
[L_q,L_p]=(q-p)L_{q+p}+\frac{c}{12}(q^3-q) \,\delta_{q+p,0},\quad q\in \mathbb Z,
\ee
where $c$ is the central charge.
Since the three generators in \eqref{Hdeform_SL2} generate the $\SL^{(q)}(2,\mathbb R)$ group, which is isomorphic to the $q$-fold cover of $\SL(2,\mathbb R)$ \cite{witten1988},
this is why we call the deformation introduced by \eqref{H_deform_intro} and \eqref{fx_sl2} the $sl(2,\mathbb R)$ deformation. The deformed Hamiltonians in \eqref{Hdeform_SL2} can be classified based on the quadratic Casimir:~\cite{ishibashi2015infinite,ishibashi2016dipolar,tada2019time}
\begin{align}\label{Eq:Casimir}
c^{(2)}&:= -(a^0)^2 + (a^+)^2 + (a^-)^2.
\end{align}
Depending on the value of $c^{(2)}$, there are three types of $sl(2,\mathbb R)$ deformed Hamiltonians as follows:
\begin{equation}
\label{3Types}
\small
\def\arraystretch{1.5}
\begin{tabular}{| l|l|l|l |}
\hline
    Quadratic Casimir  & $c^{(2)}<0$  & $c^{(2)}=0$ & $c^{(2)}>0$\\
    \hline
    Hamiltonian Type & Elliptic    & Parabolic & Hyperbolic \\
    \hline
\end{tabular}
\end{equation}
Physically, different types of Hamiltonians will give rise to different behaviors in the operator evolution \cite{wen2018floquet}. For simplicity, let us consider the ground state $|G\rangle$ of a uniform CFT for example. The system is of length $L$ with periodic boundary conditions. The path integral representation of the one-point function $\langle G|\mathcal O(w,\bar w)|G\rangle$ can be viewed as inserting a local operator in the $w$-cylinder as follows:
\be
\begin{tikzpicture}
\draw  [xshift=-40pt][thick](20pt,-40pt)--(20pt,40pt);
\draw  [xshift=-40pt][thick](70pt,-40pt)--(70pt,40pt);

\draw [xshift=-40pt][>=stealth,->] (35pt, 5pt)--(35pt,20pt);
\draw [xshift=-40pt][>=stealth,->] (35pt, 5pt)--(50pt,5pt);

\draw [xshift=-40pt](55pt, 30pt)--(55pt,40pt);
\draw [xshift=-40pt](55pt, 30pt)--(65pt,30pt);

\node at (55-40pt,5pt){$x$};
\node at (40-40pt,20pt){$\tau$};
\node at (60-40pt,35pt){$w$};

\draw [dashed](-20pt, -9.8pt)--(30pt,-9.8pt);
\node at (60-40pt,-10pt){$\bullet$};
\node at (50-40pt,-19pt){$\mathcal{O}(x,\tau)$};

\node at (20-40pt,-45pt){$x=0$};
\node at (70-40pt,-45pt){$x=L$};

\end{tikzpicture}
\ee
where $w=x+i\tau$, and the boundaries along $x=0$ and $x=L$ are identified with each other. Then the one-point function within the time dependent state $|\psi(t)\rangle=e^{-iHt}|G\rangle$ can be written as $\langle G|e^{iHt} \mathcal O(w,\bar w) e^{-iHt}|G\rangle\propto \langle G|\mathcal O(w',\bar w')|G\rangle$. Here the new location $(w',\bar w')$ of the operator is related to their original one $(w,\bar w)$ by a conformal transformation. 
To see this relation more clearly, let us first map the $w$-cylinder to a $q$-sheet Riemann surface based on the conformal mapping 
$z=e^{i\frac{2\pi q}{L}w}$, where it is reminded that $q\in \mathbb Z$ is the same as that in \eqref{Hdeform_SL2}.
On this $q$-sheet Riemann surface, one can find that 
after a time evolution with the driving Hamiltonian $H$ in \eqref{Hdeform_SL2}, the operator evolves as 
\be\label{OPevolution}
e^{i H t}\mathcal{O}(z,\bar{z})
e^{- i H t }=
\left(\frac{\partial z'}{\partial z}\right)^h
\left(\frac{\partial \bar z'}{\partial \bar z}\right)^{\bar h}
\mathcal{O}(z',\bar{z}'), 
\ee
where $z'$ is related to $z$ via a M\"obius transformation
\be
\label{z'z}
z'=\frac{\alpha z+\beta}{\beta^\ast z^\ast +\alpha^\ast}=:M\cdot z,
\ee
with $M$ being an $\SU(1,1)$ matrix  
\be
\label{matrix_M}
M=\begin{pmatrix}
\alpha &\beta\\
\beta^\ast &\alpha^\ast
\end{pmatrix}, \quad |\alpha|^2-|\beta|^2=1.
\ee
The relation between $\bar z'$ and $\bar z$ is similar if we have an anti-chiral component in the driving Hamiltonian.
Noting that $\SU(1,1)\cong \SL(2,\mathbb R)$, this is consistent with the fact that our Hamiltonians in \eqref{Hdeform_SL2} are 
$sl(2,\mathbb R)$ deformed.

The concrete forms of operator evolutions in \eqref{z'z} depend on the types of driving Hamiltonians in \eqref{3Types} as follows:\cite{wen2020periodically,han2020classification}
\begin{itemize}

\item Elliptic:
\be
\label{Elliptic}
\left\{
\begin{split}
\alpha=&\cos\left(\frac{\pi   t}{l_{\text{eff}}}\right)+i\frac{a^0}{\mathcal C}\sin\left(\frac{\pi  t}{l_{\text{eff}}}\right),\\
\beta=&i\frac{a^++ia^-}{\mathcal C}\sin\left(\frac{\pi  t}{l_{\text{eff}}}\right).
\end{split}
\right.
\ee

\item Parabolic:
\be
\label{Parabolic}
\left\{
\begin{split}
\alpha=&1+i\frac{a^0 \pi t}{l},\\
\beta=&i \frac{(a^++i a^-)\pi t}{l}.
\end{split}
\right.
\ee

\item Hyperbolic:
\be
\label{Hyperbolic}
\left\{
\begin{split}
\alpha=&\cosh\left(\frac{\pi   t}{l_{\text{eff}}}\right)+i\frac{a^0}{\mathcal C}\sinh\left(\frac{\pi t}{l_{\text{eff}}}\right),\\
\beta=&i\frac{a^++ia^-}{\mathcal C}\sinh\left(\frac{\pi  t}{l_{\text{eff}}}\right).
\end{split}
\right.
\ee
\end{itemize}
Here we have defined the effective length 
\be
\label{l_eff}
l_{\text{eff}}=\frac{l}{\mathcal C},
\ee
where $\mathcal C$ is a real number
\be
\label{Casimir_def}
\mathcal C:=\left|(a^+)^2+(a^-)^2-(a^0)^2\right|^{1/2}.
\ee
As a remark, for the driving Hamiltonian defined in \eqref{H_deform_intro} and \eqref{fx_sl2}, the chiral and anti-chiral components are not deformed independently. In this case, it is straightforward to check that the anti-chiral component in the operator evolution is determined by the same M\"obius transformation in \eqref{matrix_M},
except that now one needs to change $a^- \to -a^-$ in the expressions of $\beta$ in \eqref{Elliptic}$\sim$\eqref{Hyperbolic}.

Once we know how the operator evolves under a single Hamiltonian, it is straightforward to obtain its time evolution under multiple drivings. Suppose we drive the initial state with Hamiltonian $H_1$ for time $T_1$, then with Hamiltonian $H_2$ for time $T_2$, and so on. Then the operator evolution after $n$ steps of driving is determined by the product of a sequence of $\SU(1,1)$ matrices, with
\be
\label{z_n}
z_n=\Pi_n\cdot z,
\ee
where $\Pi_n$ acts on $z$ as defined in \eqref{z'z}, and  
\be
\label{Pi_n}
\Pi_n=M_1\cdot M_2\cdots M_n=\begin{pmatrix}
\alpha_n &\beta_n\\
\beta_n^\ast &\alpha_n^\ast
\end{pmatrix}\in \SU(1,1),
\ee
with $M_j\in \SU(1,1)$.
Now, let us introduce one main diagnostics of our quasi-periodically driven CFT: the Lyapunov exponent $\lambda_L$, which characterizes the growth of $\Pi_n$ with respect to the number of driving steps $n$, i.e., 
\be
\label{Def:Lyapunov}
\lambda_L:=\lim_{n\to \infty} \frac{|| M_1\cdot M_2\cdots M_n||}{n}
\ee
where $||\cdot ||$ is a matrix norm.
The specific choice of the matrix norm is not essential. Here we choose $|| M||:=(\sum_{j,k} |M_{j,k}|^2)^{1/2}$, where $M_{j,k}$ are the matrix elements of $M$.

The Lyapunov exponents as defined in \eqref{Def:Lyapunov} is related to the time evolution of various physical quantities, such as the entanglement entropy and energy density\cite{wen2020periodically}.
In this work, we will mainly focus on the time evolution of entanglement entropy, which serves as an ``order parameter'' to distinguish different phases as well as the phase transitions.

Now let us consider the entanglement entropy evolution after $n$ steps of driving, where the operator evolution is determined by \eqref{z_n}. To distinguish different dynamical phases, it is enough to consider a subsystem within a wavelength of deformation.
For simplicity, throughout this work we will consider the entanglement entropy evolution for the subsystem $A=[kl, (k+1)l]$ where $k\in \mathbb Z$, and it is reminded that $l$ is a single wavelength in the deformation in \eqref{fx_sl2}. Then the entanglement entropy evolution becomes\cite{wen2020periodically,2022RandomCFT}
\be
\label{S_n}
S_A(n)-S_A(n=0)=\frac{c}{3}\big(
\log|\alpha_n+\beta_n|+\log|\alpha_n'+\beta_n'|
\big),
\ee
where $\alpha_n$ and $\beta_n$ correspond to the matrix elements in $\Pi_n$ in \eqref{Pi_n} for the chiral-component driving, and similarly $\alpha'_n$ and $\beta'_n$ correspond to the contribution by the anti-chiral component.

\medskip
Note that a positive Lyapunov exponent in \eqref{Def:Lyapunov} implies that the matrix elements of $\Pi_n$ grow in $n$ as
\be
|\alpha_n|\sim |\beta_n|\sim \frac{1}{2}\, e^{\lambda_L n},
\ee
which will result in a linear growth in the entanglement entropy, i.e., $S_A(n)\propto n$. 
More explicitly, if the chiral and anti-chiral components have the same Lyapunov exponents $\lambda_L$, which is the case when the deformation is of the form in \eqref{H_deform_intro}, 
one will have $S_A(n)\simeq \frac{2\lambda_L c}{3}\cdot n$.
On the other hand, if we have a zero Lyapunov exponent, then the entanglement entropy will grow slower than linear. We will see these different features later in Sec.\ref{Sec:NoPhaseTransition} and Sec.\ref{Sec:PhaseTransition}.

\section{Quasi-periodically driven CFTs without phase transitions}
\label{Sec:NoPhaseTransition}

Let us first consider the type-I driving in Fig.\ref{Fig:Driving_Protocol}. This setup was previously studied in \cite{wen2020periodically} based on a numerical calculation in the CFT approach. It was found there is only one phase, i.e., the heating phase.
In this section, by using Avila's global theory, we show analytically why there are no phase transitions in this setup. In addition, we give an analytical expression for the Lyapunov exponents in this heating phase.

\subsection{Setup}
\label{subsec:quasiSetup1}

As shown in Fig.\ref{Fig:Driving_Protocol} (top), we consider two driving Hamiltonians $H_0$ and $H_1$, with fixed deformation parameters.
Each cycle of driving consists of two steps: We evolve the system with Hamiltonian $H_0$ first, and then with $H_1$.
In the $n$-th cycle, we evolve the system with $H_0$ for time $T_0=n\omega l$ and then with $H_1$ for time $T_1$.
If $\omega$ is a rational number $p/q$, where $p,\,q\in\mathbb Z$, then the unitary evolution generated by $e^{-iH_0T_0}$ will repeat after every $q$ cycles, and this type-I driving is reduced to a periodic driving with each period consisting of $q$ cycles. On the other hand, if $\omega$ is an irrational number, such periodicity will disappear, which gives rise to a quasi-periodic driving. Alternatively, one can certainly fix $T_0$ and vary $T_1$ quasi-periodically instead.

\medskip
Now let us give a concrete example of this quasi-periodic driving.
This example is simple enough to illustrate the main features in the type-I driving. The more general cases are studied in detail in Appendix \ref{Appendix:HeatingPhaseGerneral}. 

We consider the driving Hamiltonians $H_0$ and $H_1$ by choosing the deformation in \eqref{fx_sl2} as follows. For $H_0$ we choose 
\be
\label{H0_deform}
a^0=1, \quad a^+=-0, \quad a^-=0.
\ee
That is, $H_0$ is the uniform Hamiltonian with no deformations. For $H_1$, we can consider an arbitrary deformation in \eqref{fx_sl2}, as long as their corresponding $SU(1,1)$ matrices do not commute with each other.

\subsection{Application of Avila's global theory}

Now let us show that the above setup only gives to a heating phase with $\lambda_L>0$.

We denote the $\SU(1,1)$ matrices associated to the unitary evolutions $e^{-i H_0T_0}$ and $e^{-iH_1T_1}$ as $M_0$ and $M_1$ respectively. More concretely, we have 
\footnote{Note that here we use the convention that the factor $\pi$ is absorbed into the definition of $x$, and similarly for $\epsilon$ later in \eqref{Ax_epsilon}.}
\be
\label{Theta_0}
M_0(n)=
\begin{pmatrix}
e^{i\pi \omega\cdot n} &0\\
0 &e^{-i\pi \omega\cdot n}
\end{pmatrix}=:\begin{pmatrix}
e^{ix} &0\\
0 &e^{-ix}
\end{pmatrix}=:M_0(x),
\ee
where $n$ denotes the $n$-th cycle of driving, $\omega$ is an irrational number, and we have defined $x=\pi\omega\cdot n$. Next, $M_1$ is an arbitrary $\SU(1,1)$ matrix that does not commute with $M_0(n)$:
\be
\label{M1_general}
M_1=\begin{pmatrix}
\alpha_1 &\beta_1\\
\beta_1^\ast &\alpha_1^\ast
\end{pmatrix},
\ee
where $|\alpha_1|^2-|\beta_1|^2=1$. For this general choice, one can find that $|\alpha_1|>1$. Otherwise, $M_1$ will commute with $M_0$.

Then we consider the corresponding cocycle $(\omega, A)$, where 
\footnote{As a remark, here the degree of $M_0(x)$ is 1. This is quite different to the Schr\"odinger case \cite{Avila_global}, where the transfer matrix is homotopic to the identity, thus with zero degree. }
\be
\label{Ax}
A(x)=M_0(x)M_1.
\ee
Let us then complexify the phase by considering $A(x)\to A(x+i\epsilon)$.
Physically, this corresponds to generalizing the real time evolution in \eqref{Theta_0} to a complex time evolution.
See a related study on such generalization in \cite{2024JSM_Wen}.
Next, let $\epsilon$ in $A(x+i\epsilon)$ go to positive infinity. A direct computation yields
\be
\label{Ax_epsilon}
\begin{split}
A(x+i\epsilon)=&e^\epsilon e^{-ix}
\begin{pmatrix}
0 &0\\
0 &1
\end{pmatrix}
\begin{pmatrix}
\alpha_1 &\beta_1\\
\beta_1^\ast &\alpha_1^\ast
\end{pmatrix}+\mathcal O(1)\\
=&e^\epsilon e^{-ix}
\begin{pmatrix}
0 &0\\
\beta_1^\ast &\alpha_1^\ast
\end{pmatrix}
+\mathcal O(1).
\end{split}
\ee
Note the constant matrix $B:=\begin{pmatrix}
0 &0\\
\beta_1^\ast &\alpha_1^\ast
\end{pmatrix}$ has Lyapunov exponent
\be
\lim_{n\to \infty}\frac{\log || B^n||}{n}=
\lambda_{\text{max}}(B)=\log|\alpha_1^\ast|.
\ee
Thus we have $\lambda_L(\epsilon)=\epsilon+\log|\alpha_1^\ast|+\mathcal O(1)$. Avila's global theory tells us that as a function of $\epsilon$, $\lambda_L(\epsilon)$ is a convex and piecewise linear function, and their slopes are integers.
This implies that 
\be
\label{Eq:lambda_epsilon}
\lambda_L(\epsilon)=\text{max}\big\{\log|\alpha_1^\ast|+\epsilon, \lambda_0\big\}.
\ee
Moreover, by Avila's global theory, $(\omega, A)$ is uniformly hyperbolic if and only if $\lambda_0>0$ and $\lambda_L(\epsilon)$ is locally constant as a function of $\epsilon$, i.e., $\omega_\lambda=0$. Consequently, if $(\omega,A)$ is not uniformly hyperbolic, we have
\be
\label{Eq:lambda_0}
\lambda_L(\epsilon=0)=\text{max}\big\{\log|\alpha_1^\ast|, 0 \big\}.
\ee
In our case, as remarked below \eqref{M1_general}, we always have $|\alpha_1|=|\alpha_1^\ast|>1$. Therefore, no matter $(\omega,\,A)$ is uniformly hyperbolic or not, we always have 
\be
\lambda_L(\epsilon=0)>0.
\ee
That is, for the setup considered here, the driven CFT is always in the heating phase. This explains the numerical observation in \cite{wen2020periodically}.

Moreover, since the degree of $A(x)$ in \eqref{Ax} is 1, then the acceleration (the slope of Lyapunov exponent)  of $(\omega,A)$ is 1 if the cocycle is non-uniformly hyperbolic, i.e.,
\be
\label{omega_1}
\omega_\lambda=1. 
\ee
In the next subsection, we will show that our cocycles are indeed non-uniformly hyperbolic, with 
\be
\lambda_L(\epsilon)=\log|\alpha_1^\ast|+\epsilon,
\quad \epsilon\ge 0.
\ee
As a remark, for $\epsilon\le 0$, one will have $\lambda_L(\epsilon)=\log|\alpha_1^\ast|-\epsilon$, because $\lambda_L(\epsilon)$ is an even function of $\epsilon$ for $\SU(1,1)$ cocycles \cite{Avila_global}.
This property holds in both type-I and type-II drivings as considered in this work.

Note that if we use the same $H_0$ and $H_1$ above in the periodic driving, one can have both heating and non-heating phases \cite{han2020classification,wen2020periodically}. Interestingly, by changing to the quasiperiodic driving, only the heating phase exists.

In the above discussion, we choose a simple form of $H_0$, which corresponds to the undeformed Hamiltonian. For a general choice of elliptic $H_0$, we have the same conclusion, i.e., the type-I quasi-periodically driven system is always in the heating phase. See details in Appendix \ref{Appendix:HeatingPhaseGerneral}.

\begin{figure}
\centering
\includegraphics[width=2.5in]{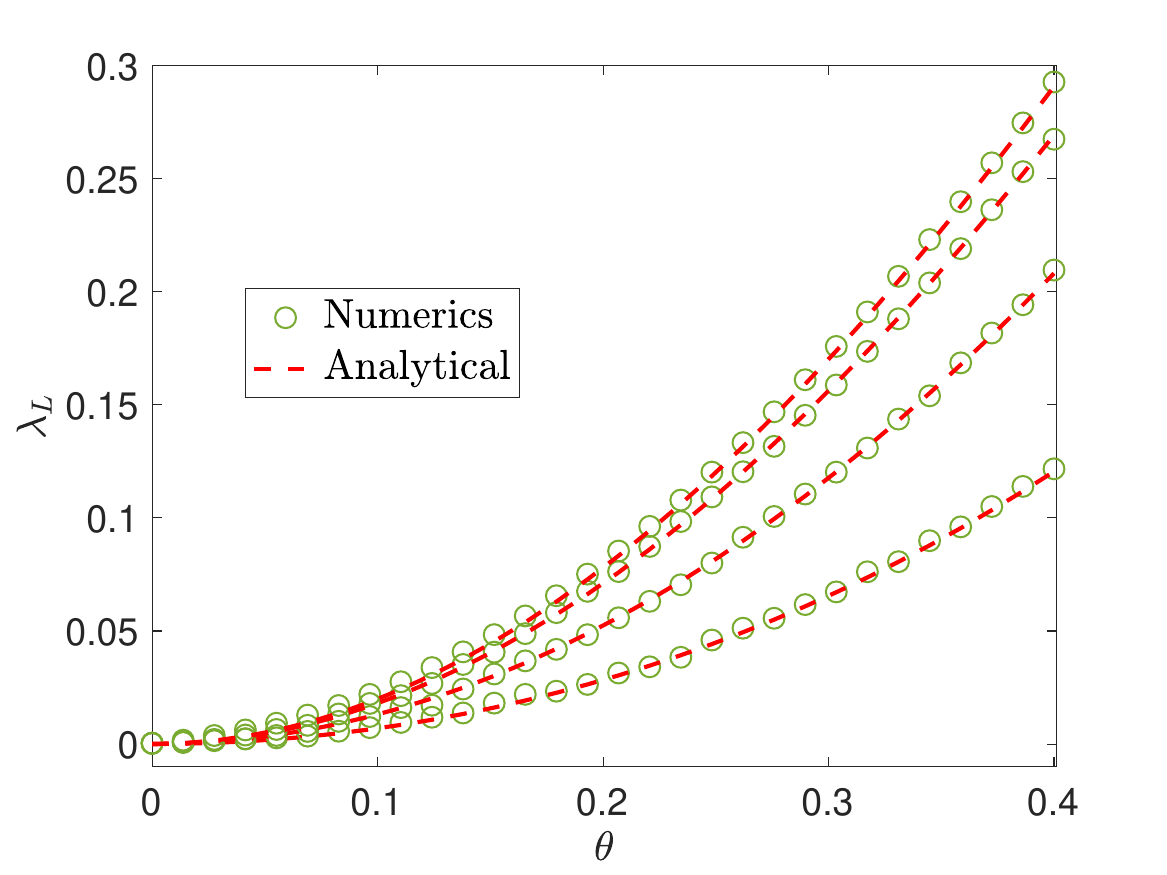}
	\caption{Lyapunov exponents $\lambda_L$ in the type-I driving in Fig.\ref{Fig:Driving_Protocol} (see also Sec.\ref{subsec:quasiSetup1}) as a function of $\theta$ that characterizes the deformation of $H_1$ in \eqref{tanh_deform}, with $q=L/l=2$.
 From bottom to top, we fix $T_1/L_{\text{eff},0}=0.2$, $0.3$, $0.4$, and $0.5$. Here $L_{\text{eff},0}=L\cosh(2\theta_0)$
 with $\theta_0=0.1$. The numerical results are obtained from 
 \eqref{Def:Lyapunov}, \eqref{Theta_0}, and \eqref{M1_theta_maintext}, and we choose $\omega=(\sqrt 5-1)/2$ in \eqref{Theta_0}. The analytical results are obtained from \eqref{lambda_epsilon_example} by setting $\epsilon=0$.
	}
 \label{Fig:LyapunovHeating}
\end{figure}

\subsection{Lyapunov exponents and acceleration}

Let us give a concrete example to illustrate the complexified Lyapunov exponents and acceleration obtained in the previous subsection.

For the driving Hamiltonian $H_1$, we consider the deformation in \eqref{fx_sl2} with
\be
\label{tanh_deform}
a^0=1, \quad a^+=-\tanh(2\theta), \quad a^-=0,
\ee
where $\theta\in(0,+\infty)$.
Based on \eqref{Elliptic}, the expression for $M_1$ is  
\be
\label{M1_theta_maintext}
    M_1 = \begin{pmatrix}
    \alpha_1 & \beta_1 \\ \beta_1^* & \alpha_1^*
    \end{pmatrix},
\ee
where 
\be
\label{M1_alpha_beta}
	\left\{
	\begin{split}
		&\alpha_1=\cos{\left( \frac{\pi T_1}{l_{\text{eff}}} \right)} + i\cosh(2\theta)\cdot \sin{\left( \frac{\pi T_1 }{l_{\text{eff}}} \right)},\\
		&\beta_1=- i\sinh(2\theta)\cdot\sin{\left( \frac{\pi T_1}{l_{\text{eff}}} \right)}.
	\end{split}
	\right.
\ee
Here we have defined $l_{\text{eff}}=l \cosh(2\theta)$ according to the definition in \eqref{l_eff}. During the quasiperiodic driving, $T_1$ is fixed in \eqref{M1_alpha_beta}.

Then based on our result in \eqref{Eq:lambda_epsilon}, if the cocycle is not uniformly hyperbolic, we will have 
\be
\label{lambda_epsilon_example}
\lambda_L(\epsilon)=\epsilon+\frac{1}{2}\log\left[
1+\sinh^2(2\theta)\cdot \sin^2\Big(\frac{\pi T_1}{l_{\text{eff}}}\Big)
\right].
\ee
As shown in Fig.\ref{Fig:LyapunovHeating}, we compare the numerical results of Lyapunov exponents $\lambda_L(\epsilon=0)$ obtained from \eqref{Def:Lyapunov} and the analytical results in \eqref{lambda_epsilon_example} by setting $\epsilon=0$. They agree with each other perfectly. 

\begin{figure}
\centering
\includegraphics[width=2.5in]{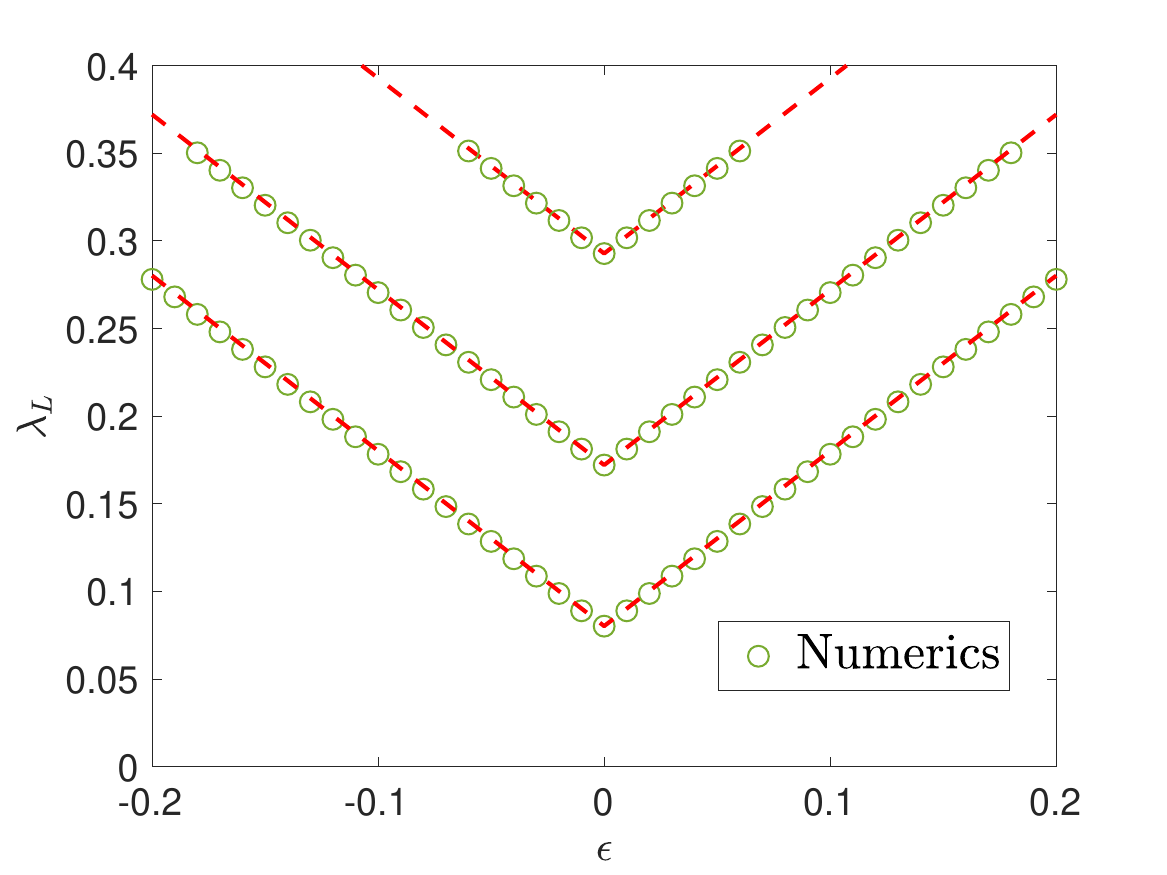}
	\caption{Complexified Lyapunov exponents $\lambda_L(\epsilon)$ in \eqref{Complex_Lyapunov_0} as a function of $\epsilon$ in the type-I quasiperiodic driving in Fig.\ref{Fig:Driving_Protocol}.
We fix $T_1/L_{\text{eff},0}=0.5$, and choose $\theta=0.2$, $0.3$, and $0.4$ in \eqref{tanh_deform} from bottom to top.
Here $L_{\text{eff},0}=L\cosh(2\theta_0)$
 with $\theta_0=0.1$. The dashed red lines correspond to $\lambda_L(\epsilon)=\lambda_L(\epsilon=0)+|\epsilon|$.
 The features here correspond to the supercritical case
 in Fig.\ref{Lyapunov_Phase}.
}
 \label{Fig:LyapunovHeating_Acc}
\end{figure}

In addition, we checked the complexified Lyapunov exponents $\lambda_L(\epsilon)$ in Fig.\ref{Fig:LyapunovHeating_Acc}, which give $\omega_\lambda=1$
and agree with the result in \eqref{omega_1}.
Based on the agreement in Fig.\ref{Fig:LyapunovHeating} and Fig.\ref{Fig:LyapunovHeating_Acc}, it is confirmed that the cocycles we consider here are non-uniformly hyperbolic, and the corresponding Lyapunov exponents are given by $\lambda_L(\epsilon)=\log|\alpha_1^\ast|+\epsilon$.

\subsection{Entanglement entropy evolution}
\label{subSec:EE_evolution}

Now let us verify the CFT prediction based on a lattice free fermion calculation, by calculating the entanglement entropy evolution. For $H_0$, we choose $(a^0,\,a^+,\,a^-)=(1,\, 0,\, 0)$. For $H_1$, we choose the parametrization in \eqref{tanh_deform}.

\begin{figure}
\centering
\includegraphics[width=2.5in]{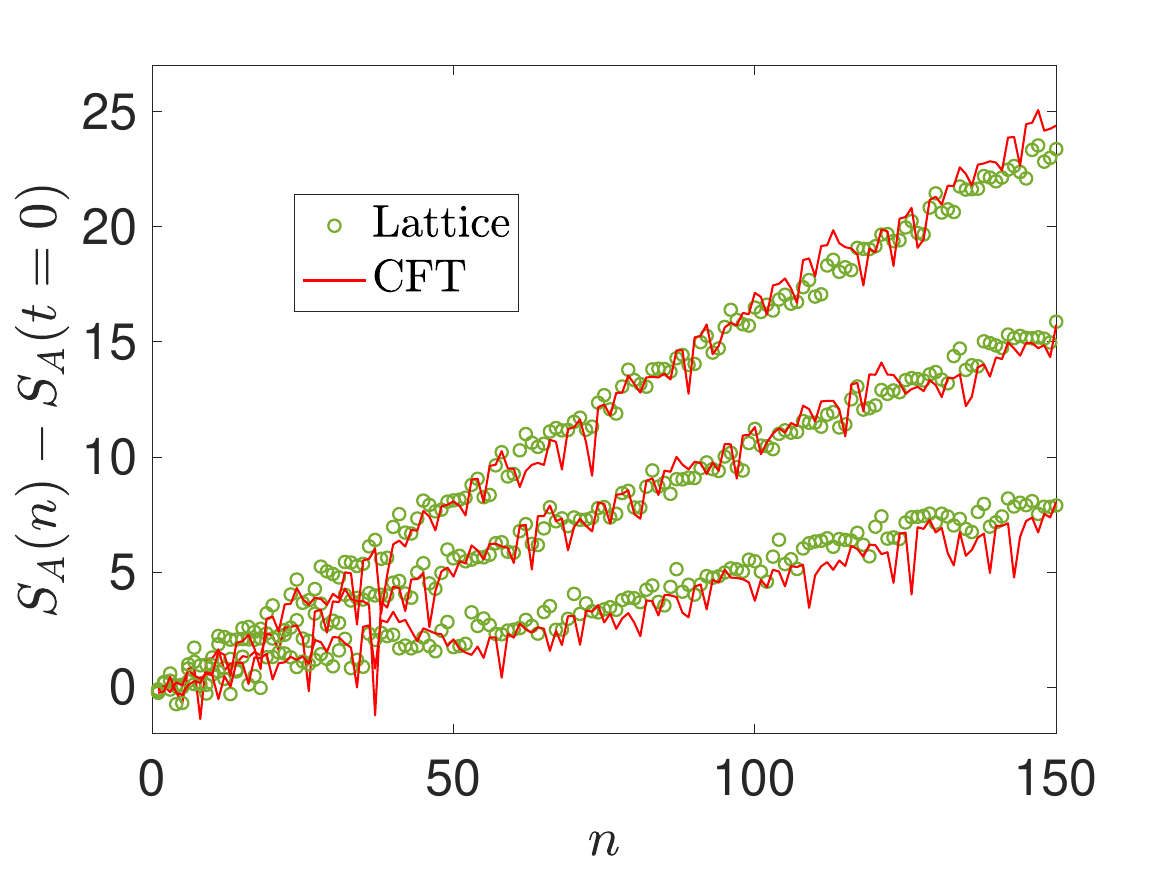}
	\caption{Comparison of entanglement entropy evolution for lattice-model and CFT calculations for the type-I quasiperiodic driving in Fig.\ref{Fig:Driving_Protocol}. From top to bottom, we choose $\theta=0.45$, $0.4$ and $0.35$ in the deformation of $H_1$ in \eqref{tanh_deform}, and the other parameters are the same as those in Fig.\ref{Fig:LyapunovHeating_Acc}.
    The CFT result is based on \eqref{S_n}, \eqref{Theta_0}, and \eqref{M1_theta_maintext}.
    Here we choose $L=600$ with periodic boundary conditions and the subsystem is $A=[0,L/2]$.
}
 \label{Fig:EE_NoPhaseTransition}
\end{figure}

To compare with our CFT results, we will consider a lattice free fermion model, the low energy physics of which is described by a $c=1$ free Dirac fermion CFT.
We choose the initial state as the ground state of $H_0$ with half-filling, where
\be
H_0=-\frac{1}{2}\sum_{j=1}^{L}c_j^\dag c_{j+1}+h.c.,
\ee
with periodic boundary conditions.
The fermionic operators $c_j$ satisfy the anti-commutation relation 
$\{c_i,c_j^\dag\}=\delta_{ij}$, and $\{c_i, c_j\}=\{c_i^\dag, c_j^\dag\}=0$. The $sl(2,\mathbb R)$  deformed Hamiltonian can be constructed as
\be
\label{H1_lattice}
H_1=-\frac{1}{2}\sum_{j=1}^{L} f(j)\, c_j^\dag c_{j+1}+h.c.,
\ee
where $f(j)$ is the discrete version of $f(x)$ in \eqref{fx_sl2} with the values of $(a^0,\, a^+, \,a^-)$ chosen in \eqref{tanh_deform}, and $q=L/l=2$.
Then one can calculate the entanglement entropy evolution based on the standard procedure by using two-point correlation matrix \cite{2003_Peschel}.
A sample plot of the entanglement entropy evolution for subsystem $A=[0,l]$ is shown in Fig.\ref{Fig:EE_NoPhaseTransition}, where one can observe a linear growth of entanglement entropy in both lattice and CFT calculations, as expected.

\section{Quasi-periodically driven CFTs with phase transitions}
\label{Sec:PhaseTransition}

In this section, we will consider the type-II quasiperiodic driving as illustrated in Fig.\ref{Fig:Driving_Protocol} (bottom). In this new setup, we find there are phase transitions during the quasiperiodic driving, which can be analytically studied.

\subsection{Setup}
\label{subsec:quasiSetup2}

As shown in Fig.\ref{Fig:Driving_Protocol} (bottom), in the type-II quasiperiodic driving, we fix the driving time $T_0$ and $T_1$ in each driving cycle, but change the driving Hamiltonians quasi-periodically in time.
In the following, let us illustrate with a concrete example, which give us all the interesting features that will appear in the general cases. One can refer to Appendix \ref{Appendix:PhaseTransition} for the general discussions.

More concretely, we consider the choices that $H_0$ are hyperbolic. The driving Hamiltonians $H_0$ depend on time in a quasiperiodic way by choosing the deformation in \eqref{fx_sl2} with
\be
a^0=0,\quad  a^+= \cos(\pi \omega\cdot n),\quad  a^-=\sin(\pi \omega\cdot n),
\ee
where $\omega$ is an irrational number, and $n\in \mathbb Z$ denotes the $n$-th driving cycle. One can find $\mathcal C=1$ in the definition in \eqref{Casimir_def}.
It is reminded that each driving cycle consists of two steps: we drive the system with Hamiltonian $H_0$ for time $T_0$, and then with Hamiltonian $H_1$ for time $T_1$.

With the above choice of $H_0$, based on \eqref{Hyperbolic}, the corresponding $\SU(1,1)$ matrix that determines the operator evolution in the $n$-th cycle is
\be
\label{M0_PT_maintext}
M_0(n)=\begin{pmatrix}
\cosh(\alpha) &ie^{i\pi\omega\cdot n}\sinh(\alpha)\\
-ie^{-i\pi\omega\cdot n}\sinh(\alpha) &\cosh(\alpha)
\end{pmatrix},
\ee
where we have defined $\alpha:=\pi T_0/l>0$.

 For $H_1$, we can consider an arbitrary deformation in \eqref{fx_sl2}, as long as the corresponding $\SU(1,1)$ matrix does not commute with $M_0$, i.e., $[M_0,M_1]\neq 0$.


\begin{figure}[tp]
\centering
\begin{tikzpicture}
  \node[inner sep=0pt] (russell) at (-65pt,-20pt)
    {\includegraphics[width=3.0in]{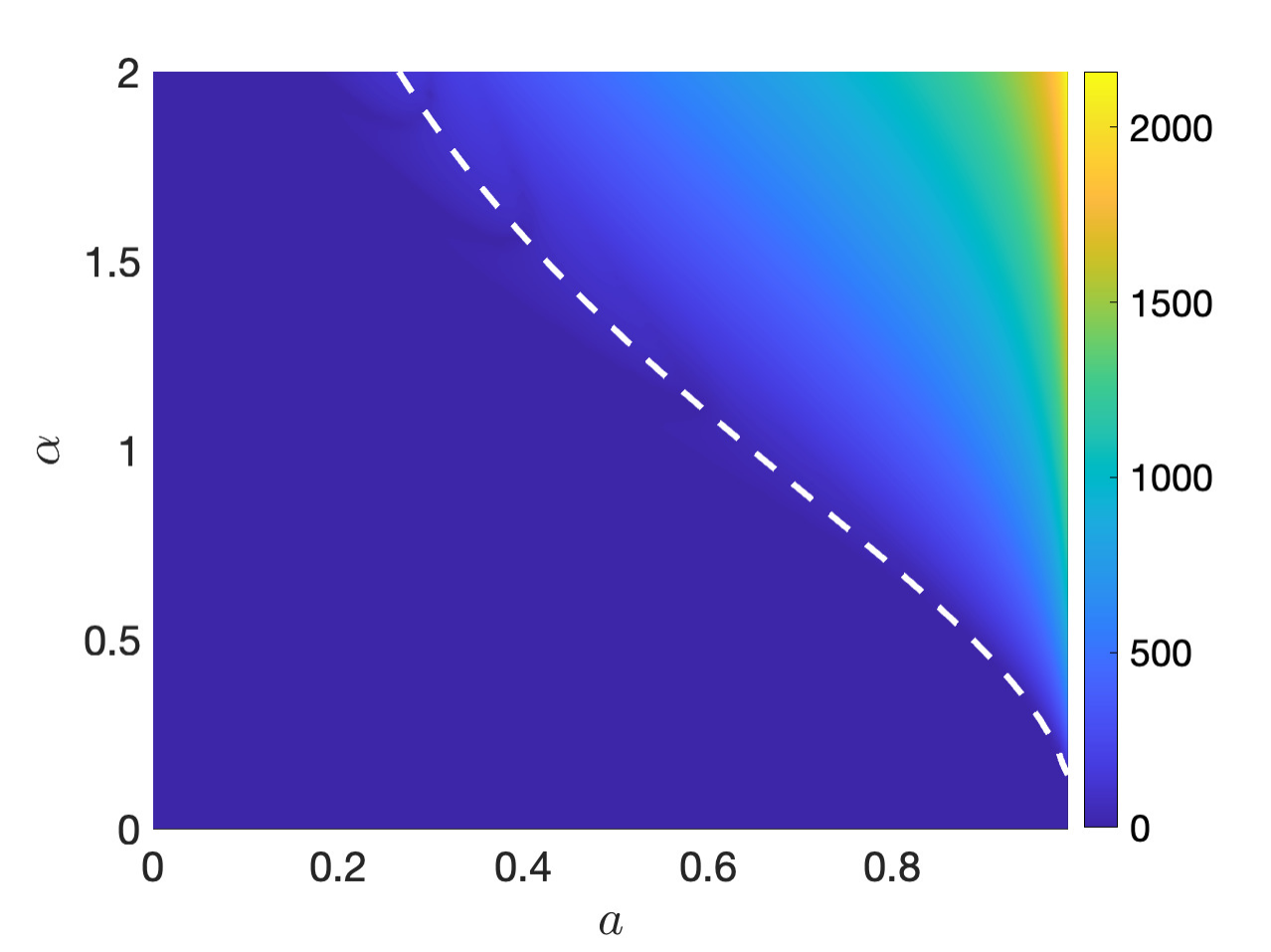}};
  \normalsize
 \node at (-100pt, -30pt){\textcolor{white}{Non-heating}};
 \node at (-35pt, +20pt){\textcolor{white}{Heating}};                   
    \end{tikzpicture}
    \caption{
Phase diagram of a quasi-periodically driven CFT as a function of the driving parameters $\alpha$ and $a$ in \eqref{M0_PT_maintext} and \eqref{M1_3}, respectively. The quasiperiodic frequency in \eqref{M0_PT_maintext} is set to be $\omega=\pi(\sqrt{5}-1)/2$.
The white dashed line corresponds to the analytical result of phase transitions in \eqref{PhaseTransition_1}.
The numerical values correspond to the entanglement entropy of $A=[0,\,l]$ after $N=1000$ cycles of driving. 
}
 \label{Fig:Phase}
\end{figure}

\subsection{Application of Avila's global theory}

Based on the setup introduced above, let us consider the cocyle $(\omega, A)$ where 
\be
\begin{split}
A(x)=&M_0(x)M_1\\
=&
\begin{pmatrix}
\cosh(\alpha) &ie^{ix}\sinh(\alpha)\\
-ie^{-ix}\sinh(\alpha) &\cosh(\alpha)
\end{pmatrix}
\begin{pmatrix}
\alpha_1 &\beta_1\\
\beta_1^\ast &\alpha_1^\ast
\end{pmatrix}.
\end{split}
\ee
Here we have defined $x:=\pi\omega\cdot n$, and $M_1$ is a general $\SU(1,1)$ matrix.

Let us then complexify the phase by replacing $x$ with $x+i\epsilon$.
Physically, this means we need to generalize the driving Hamiltonians from Hermitian to non-Hermitian.
Next, let us take the limit $\epsilon\to +\infty$. Then one can obtain
\be
\begin{split}
A(x+i\epsilon)
=&e^\epsilon e^{-ix}
\begin{pmatrix}
0 &0\\
-i\sinh(\alpha) &0
\end{pmatrix}
\begin{pmatrix}
\alpha_1 &\beta_1\\
\beta_1^\ast &\alpha_1^\ast
\end{pmatrix}+\mathcal O(1)\\
=&
-i \sinh(\alpha)\,  e^\epsilon e^{-ix}
\begin{pmatrix}
0 &0\\
\alpha_1 &\beta_1
\end{pmatrix}
+\mathcal O(1).
\end{split}\nonumber
\ee
The constant matrix that appears above, i.e., $B=-i\sinh(\alpha)\begin{pmatrix}
0 &0\\
\alpha_1 &\beta_1
\end{pmatrix}$, has the Lyapunov exponent
\be
\lim_{n\to \infty}\frac{\log||B^n ||}{n}=\lambda_{\text{max}}(B)=
\log\left|\beta_1 \cdot \sinh(\alpha)\right|.
\nonumber
\ee
Therefore, we have $\lambda_\epsilon(A)=\epsilon+\log\left|\beta_1 \cdot \sinh(\alpha)\right|+\mathcal O(1)$. Avila's theory shows that as a function of $\epsilon$, $\lambda_\epsilon(A)$ is a convex and linear function, and their slopes are integers. 
This implies that
\be
\label{lambda_L_Avila_maintext0}
\lambda_\epsilon(A)=\max\left\{
\log\left|\beta_1 \cdot \sinh(\alpha)\right|+\epsilon, 
\lambda_0(A)
\right\}.
\ee
Moreover, by Avila's global theory, $(\omega,A)$ is uniformly hyperbolic if and only if $\lambda_0(A)>0$ and the acceleration $\omega_\lambda=0$.
Consequently, if $(\omega, A)$ is not uniformly hyperbolic, it will correspond to one of the three cases in Fig.\ref{Lyapunov_Phase}, and the corresponding Lyapunov exponent has the expression:
\be
\label{lambda_L_Avila_maintext0}
\lambda_L(\epsilon)=\max\left\{
\log\left|\beta_1 \cdot \sinh(\alpha)\right|+\epsilon, 
0
\right\}.
\ee
Different from the type-I driving in Sec.\ref{Sec:NoPhaseTransition}, here $\lambda_{L}(\epsilon=0)$ may be zero, indicating the possible phase transitions.  
More explicitly, the heating phase is determined by $\lambda_L(\epsilon=0)>0$, i.e., 
$
|\beta_1\cdot \sinh(\alpha)|>1,
$
and the phase transitions to the non-heating phases are determined by
\be
\label{PhaseTransition_0}
\left|\beta_1\cdot \sinh(\alpha)\right|= 1.
\ee
In the next subsection, we will give a concrete example to illustrate such phase transitions.

\begin{figure}[tp]
\centering
\begin{tikzpicture}
  \node[inner sep=0pt] (russell) at (-65pt,-20pt)
    {\includegraphics[width=2.8in]{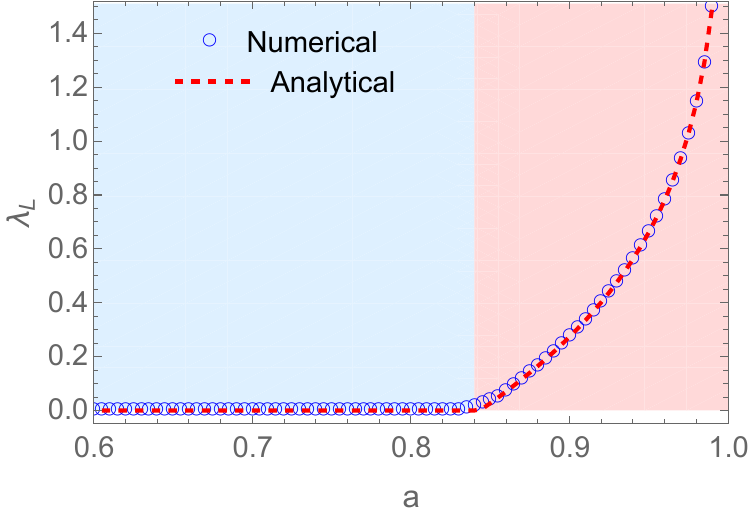}};
  
 \node at (-95pt, -10pt){\textcolor{blue}{Non-heating}};
 \node at (-15pt, -10pt){\textcolor{red}{Heating}};                   
    \end{tikzpicture}
    \caption{Comparison of Lyapunov exponents obtained from
    numerical calculations based on \eqref{Def:Lyapunov} and the analytic formula in \eqref{lambda_L_Avila_maintext}. We choose $\alpha = 0.6$ in \eqref{M0_PT_maintext}.  
    The phase transition point is determined by \eqref{PhaseTransition_1}.
    }
 \label{Fig:Lypidx_Compare_Quasi}
\end{figure}

\subsection{Lyapunov exponents and accelerations}

We consider a concrete choice of $H_1$, with the following simple deformation
\be
a^0=1, \quad a^+=a, \quad a^-=0,
\ee
where $a\in\mathbb R$ and $|a|<1$. Based on \eqref{Elliptic}, one can find the corresponding $\SU(1,1)$ matrix is
\be
\label{M1_3}
M_1=\frac{i}{\mathcal C}
\begin{pmatrix}
1  &a\\
-a &-1
\end{pmatrix}, \quad \mathcal{C}=\sqrt{1-a^2}.
\ee
Then based on our result in \eqref{lambda_L_Avila_maintext0}, if the cocycle is not uniformly hyperbolic, we will have
\be
\label{lambda_L_Avila_maintext}
\lambda_L(\epsilon)=\max\left\{
\log\left|\frac{a}{\sqrt{1-a^2}} \cdot \sinh(\alpha)\right|+\epsilon, 
0
\right\}.
\ee
In the two dimensional parameter space spanned by $a$ and $\alpha$, the heating phase is determined by $\lambda_L(\epsilon=0)>0$, and the phase transition is determined by
\be
\label{PhaseTransition_1}
\left|\frac{a}{\sqrt{1-a^2}} \cdot \sinh(\alpha)\right|= 1,
\ee
which corresponds to the white dashed line in Fig.\ref{Fig:Phase}.
One can find the phase diagram obtained from the analytical results agrees very well with the distinct features of entanglement entropy evolution in different phases.

We further compare the Lyapunov exponents that are obtained from the analytical formula in \eqref{lambda_L_Avila_maintext} and those obtained from the numerical results based on \eqref{Def:Lyapunov}. 
As shown in Fig.\ref{Fig:Lypidx_Compare_Quasi}, 
which corresponds to a fixed $\alpha$ in Fig.\ref{Fig:Phase},
the numerical and analytical results for $\lambda_L(\epsilon=0)$ agree very well for a large range of parameters.

\begin{figure}[tp]
\centering
\begin{tikzpicture}
  \node[inner sep=0pt] (russell) at (0pt,-20pt)
    {\includegraphics[width=2.2in]{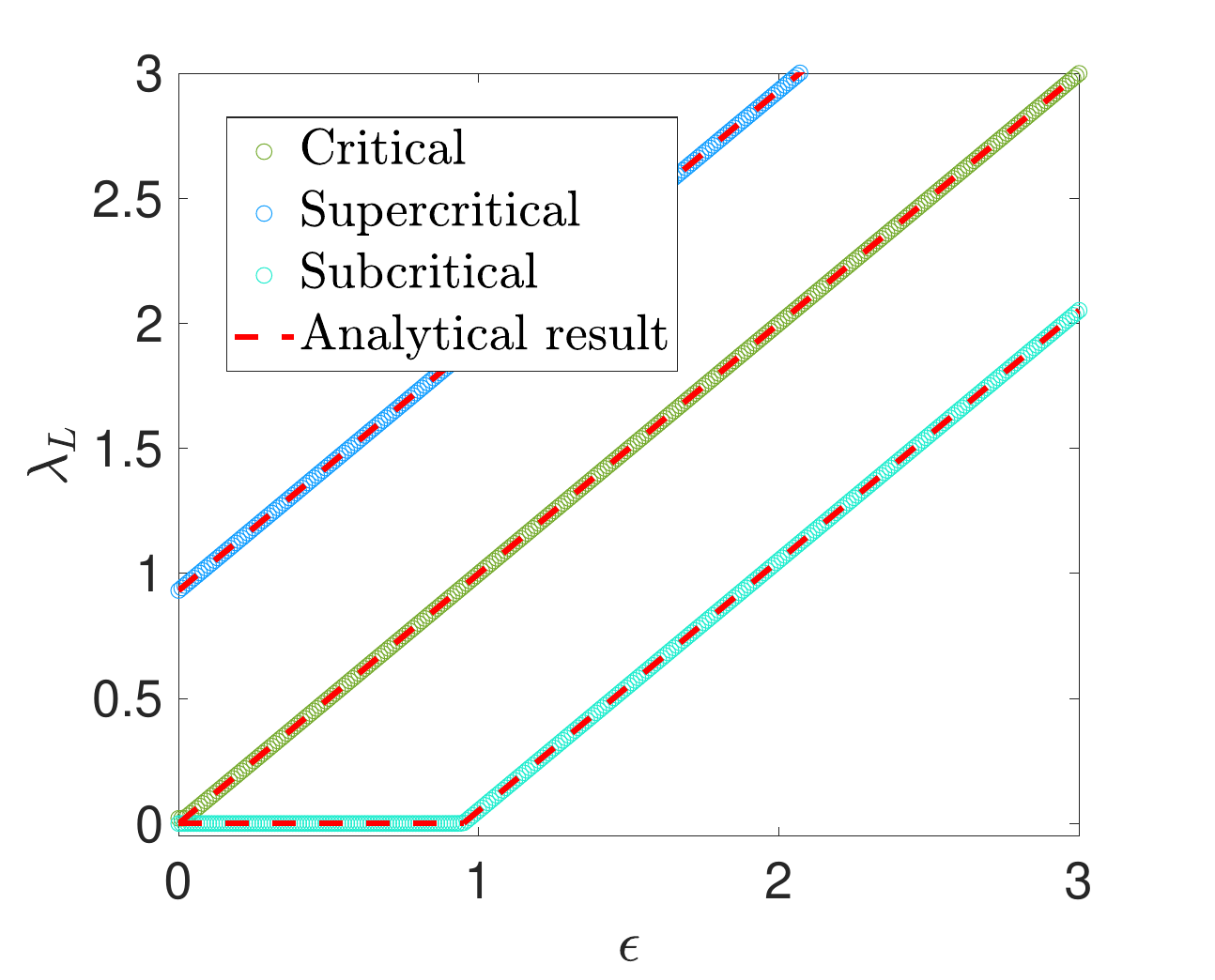}};

      \node[inner sep=0pt] (russell) at (0pt,-148pt)
    {\includegraphics[width=2.2in]{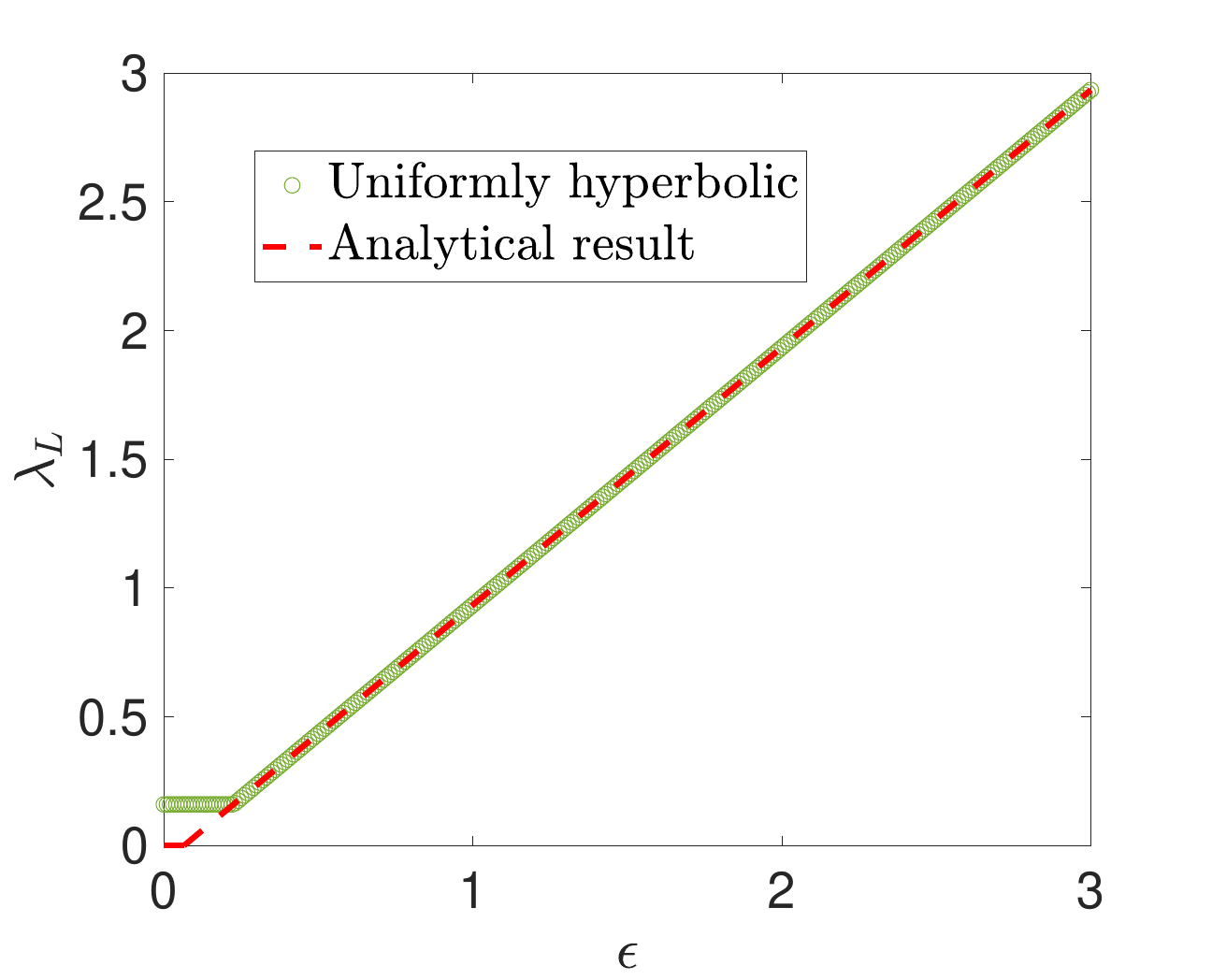}};
  
    \end{tikzpicture}
    \caption{
    Different behaviors of complexified Lyapunov exponents $\lambda_L(\epsilon)$ as a function of $\epsilon$ in our type-II
    quasiperiodic driving in Fig.\ref{Fig:Driving_Protocol}. See details of the driving in Sec.\ref{subsec:quasiSetup2}.
    The dots correspond to numerical results based on \eqref{Def:Lyapunov}, and the red dashed lines are analytical results in \eqref{lambda_L_Avila_maintext}, which assume the cocycles are not uniformly hyperbolic.
    One can observe different features corresponding to the 
    critical (phase transition), supercritical (heating), subcritical (non-heating), and 
    the uniform-hyperbolic (heating) cases.    
    It is noted that $\lambda_L(\epsilon)$ is an even function of $\epsilon$, and therefore the region for $\epsilon<0$ is not shown here. The parameters we choose are $(a,\,\alpha)=(0.8417,0.6041)$ for the critical (phase transition) case , $(0.8,\,1.4)$ for the supercritical (heating) case, $(0.4,0.8)$ for the subcritical (non-heating) case, and $(0.25,2)$ for the uniformly hyperbolic (heating) case.
    }
 \label{Fig:Lypidx_Epsl_Quasi}
\end{figure}

We want to emphasize that, however, there could be some subtle deviations between \eqref{lambda_L_Avila_maintext} and the numerical calculations at certain driving parameters near the phase transitions, where it is observed that the cocycles may become uniformly hyperbolic. In this case, the Lypunov exponents are no longer described by \eqref{lambda_L_Avila_maintext} which only works for the cases that are not uniformly hyperbolic. Such deviations can be seen clearly by studying the complexified Lyapunov exponents, as shown in Fig.\ref{Fig:Lypidx_Epsl_Quasi}.
It is observed that when the cocycles are not uniformly hyperbolic, the numerical results  agree with the analytical result in \eqref{lambda_L_Avila_maintext} very well, as seen in Fig.\ref{Fig:Lypidx_Epsl_Quasi} (top). However, when the cocycles become uniformly hyperbolic with $\lambda_L(\epsilon=0)>0$ and $\omega_\lambda=0$, the numerical results will go beyond the scope of \eqref{lambda_L_Avila_maintext}, as shown in Fig.\ref{Fig:Lypidx_Epsl_Quasi} (bottom), where \eqref{lambda_L_Avila_maintext} gives a subcritical behavior.

\subsection{Entanglement entropy evolution}

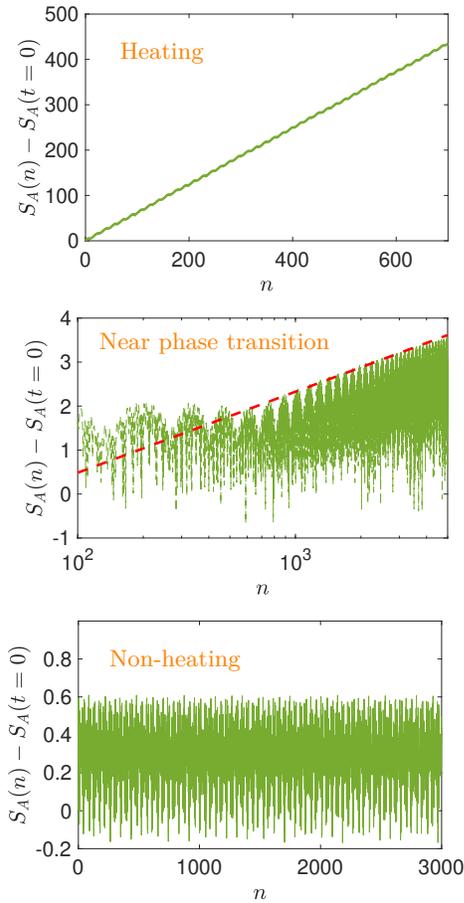
\begin{figure}[tp]
\centering
\begin{tikzpicture}
  \node[inner sep=0pt] (russell) at (-65pt,-20pt)
    {\includegraphics[width=2.5in]{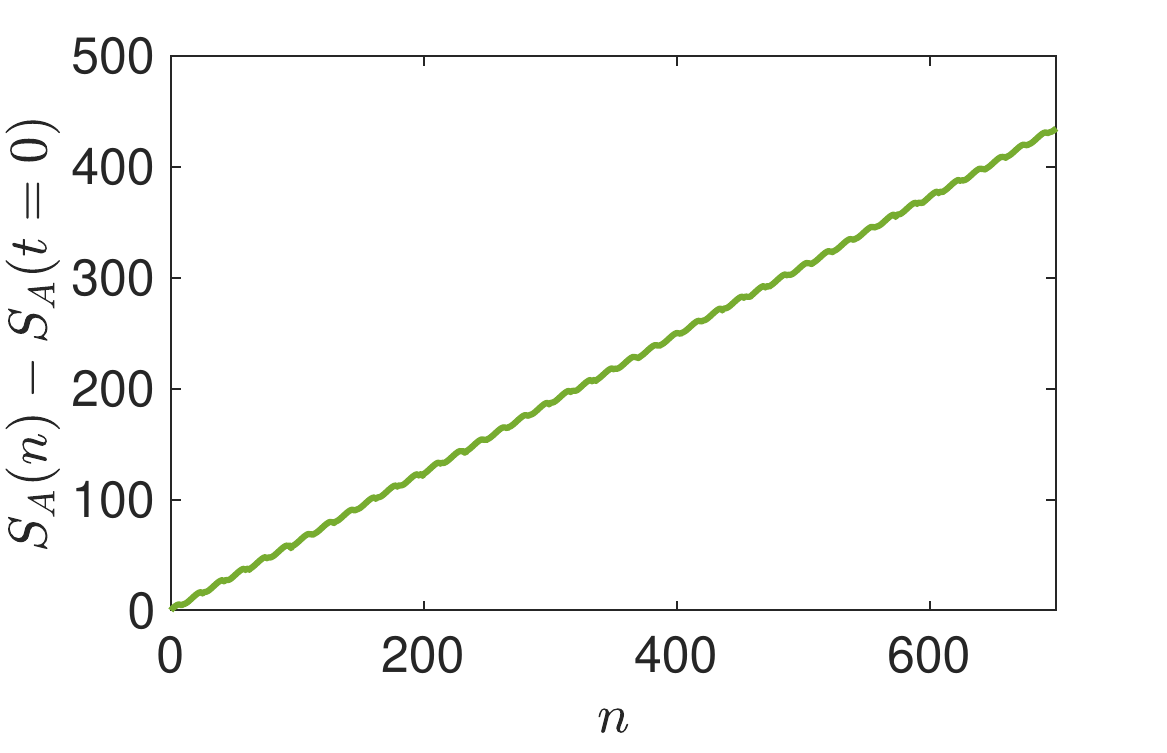}};

      \node[inner sep=0pt] (russell) at (-65pt,-135pt)
    {\includegraphics[width=2.5in]{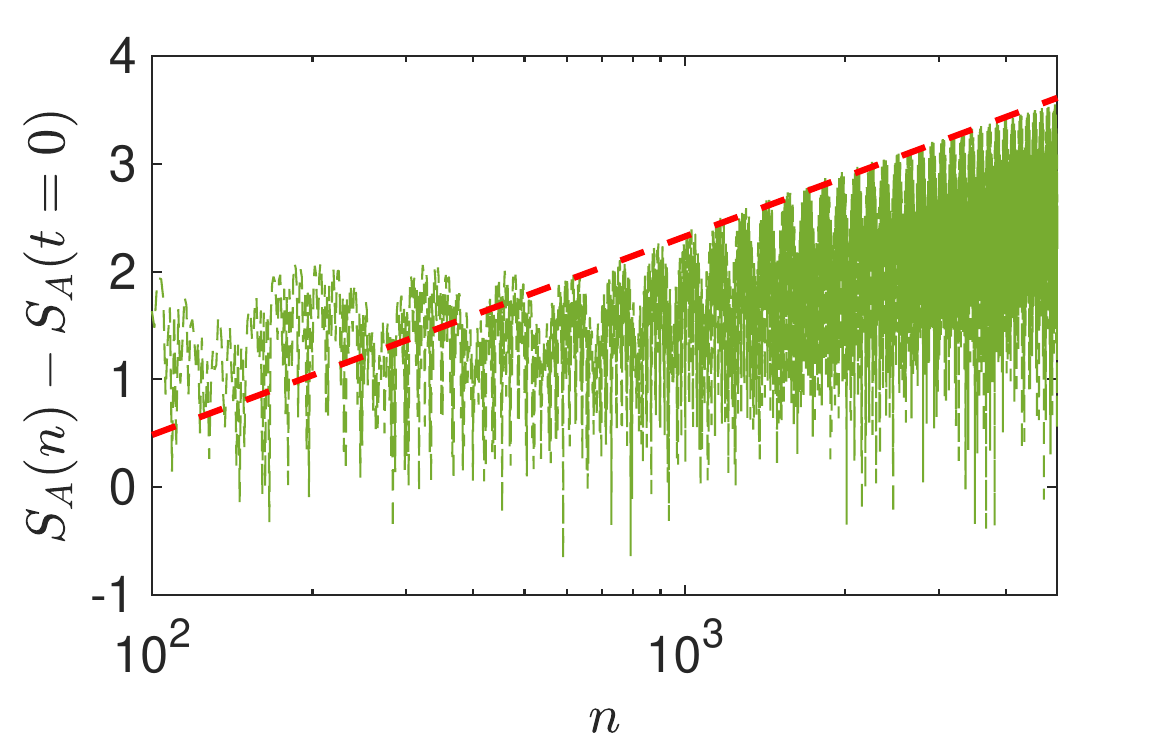}};

      \node[inner sep=0pt] (russell) at (-68pt,-250pt)
    {\includegraphics[width=2.52in]{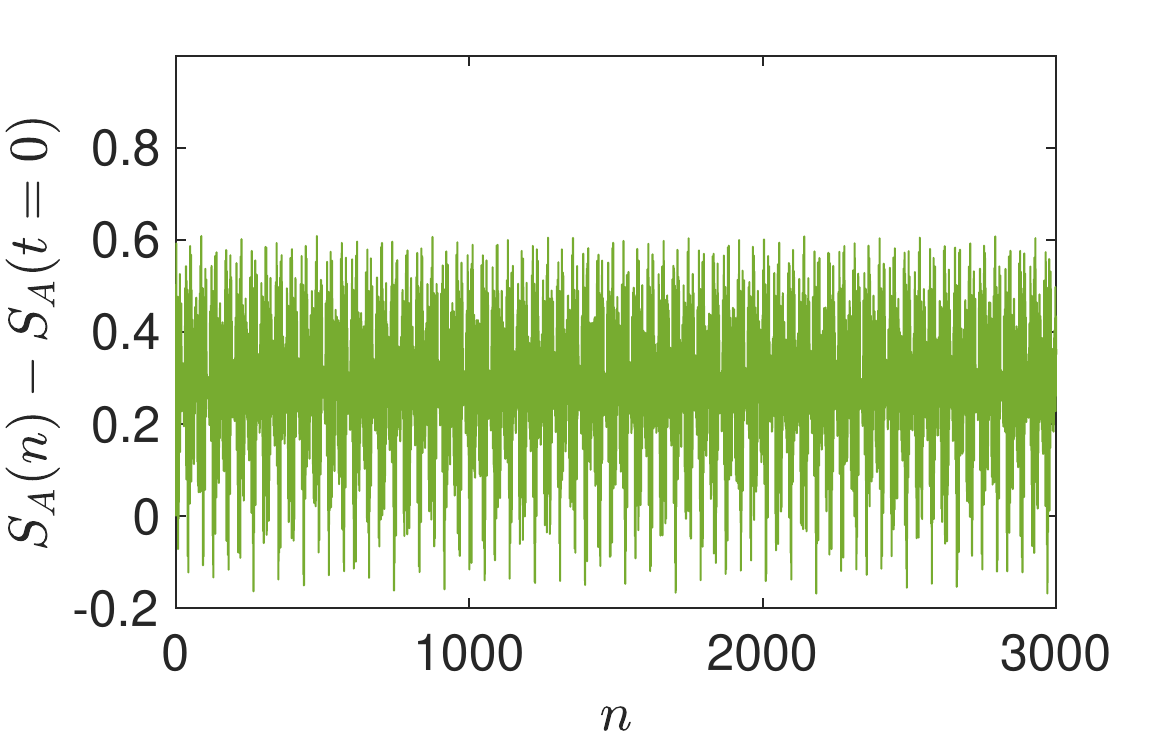}};

 \node at (-100pt, 15pt){\textcolor{orange}{Heating}};
 \node at (-80pt, -95pt){\textcolor{orange}{Near phase transition}};
 \node at (-95pt, -215pt){\textcolor{orange}{Non-heating}};

    \end{tikzpicture}
    \caption{From top to bottom: CFT calculations of entanglement entropy evolution for $A=[0,l]$ in the heating phase, near the phase transition, and in the non-heating phase. The red dashed line is a guiding line of the form $y=a\log(n)+b$. The parameters we choose here 
    are $(a,\alpha)=(0.8,1.4)$, $(0.832926,0.6)$, and $(0.4,0.8)$, respectively.}
 \label{Fig:Lypidx_Epsl_Quasi}
\end{figure}

Now, let us consider the time evolution of entanglement entropy during the type-II quasiperiodic driving, based on both CFT and lattice-model calculations.

Similar to what we did in Sec.\ref{subSec:EE_evolution}, 
We consider the entanglement entropy evolution for a subsystem $A=[0,l]$, where it is reminded that $l$ is the wavelength of deformation in \eqref{fx_sl2}. 

As shown in Fig.\ref{Fig:Lypidx_Epsl_Quasi} are the entanglement entropy evolutions in different phases and near the phase transition, which are obtained based on the CFT results in \eqref{S_n}. The entanglement entropy grows linearly the heating phase, and oscillates in time in the non-heating phase. Near the phase transition \footnote{Note that in numerics it is hard to pin down the \textit{exact} location of the phase transition, since near phase transitions there could be uniformly hyperbolic cases with very small Lyapunov exponents, as discussed in the previous subsection. Therefore, rigorously speaking, here we can only study the point \textit{near} the phase transition in numerics.
}, it is observed that the entanglement entropy grows logarithmically in time, i.e., $S_A(n)\propto \log (n)$.
Note that such $\log(n)$ growth behavior was also observed at the phase transition in a periodically driven CFT \cite{wen2020periodically,wen2018floquet}.

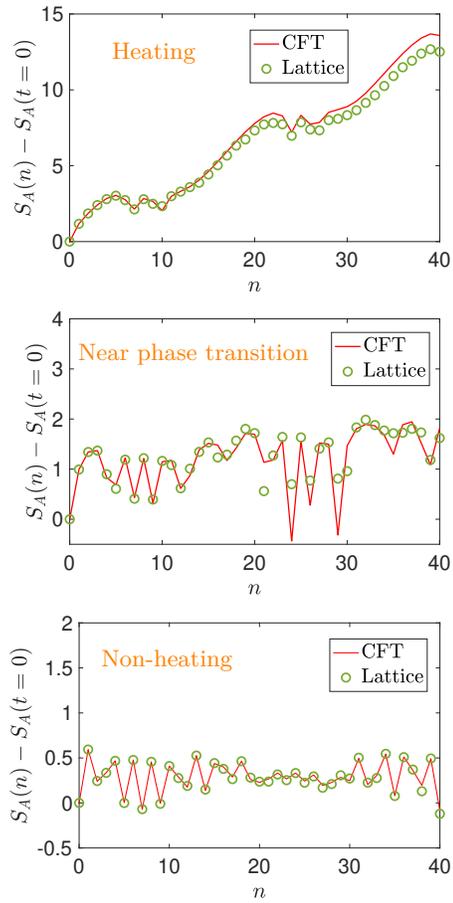
\begin{figure}[tp]
\centering
\begin{tikzpicture}
  \node[inner sep=0pt] (russell) at (-65pt,-20pt)
    {\includegraphics[width=2.5in]{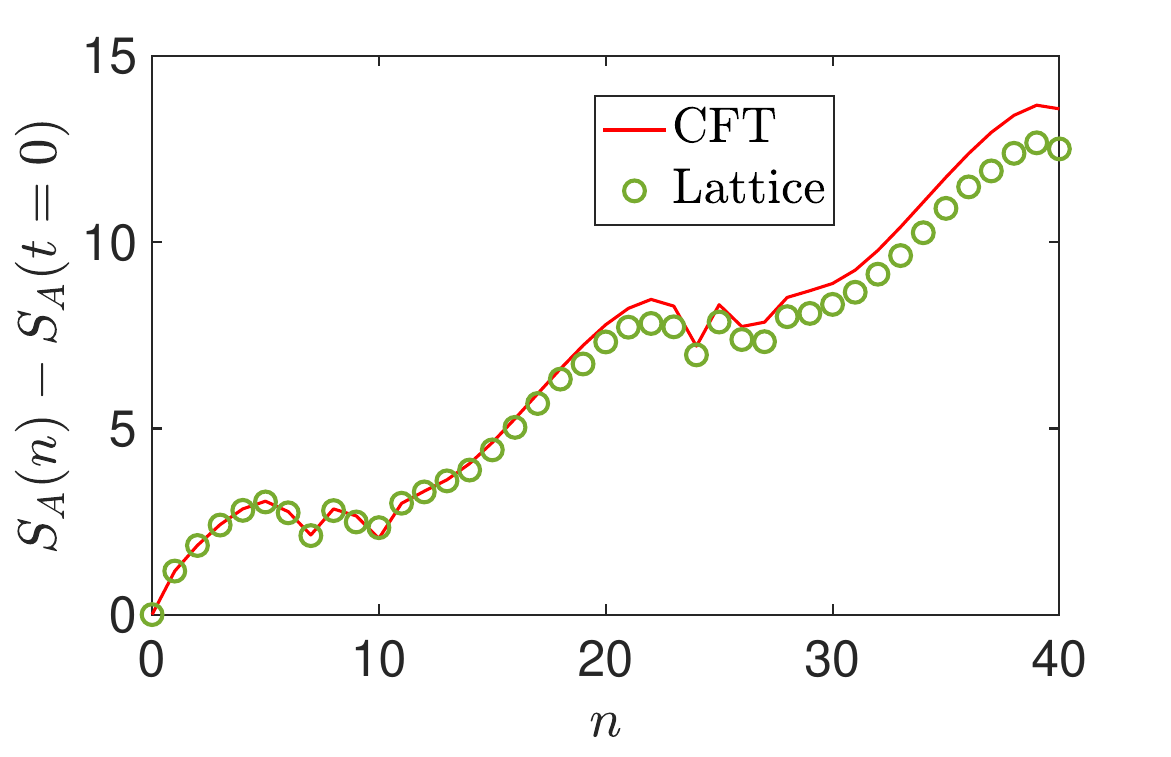}};

          \node[inner sep=0pt] (russell) at (-65pt,-135pt)
    {\includegraphics[width=2.5in]{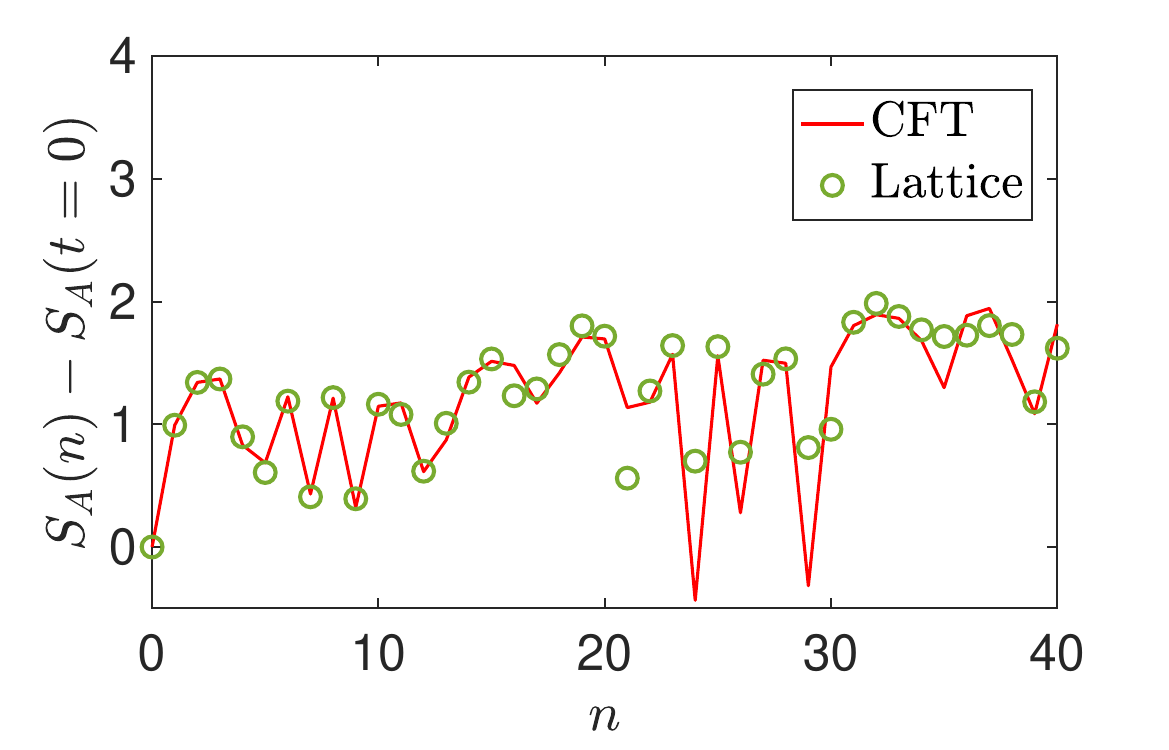}};

          \node[inner sep=0pt] (russell) at (-65pt,-250pt)
    {\includegraphics[width=2.5in]{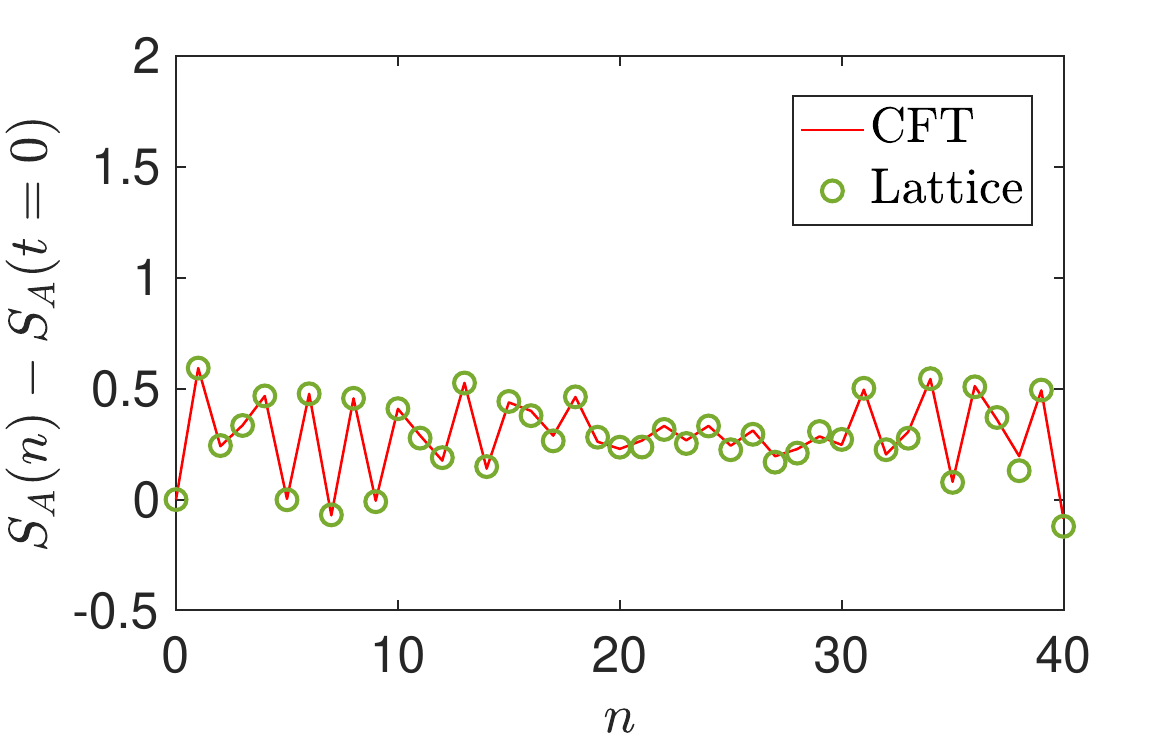}};
  
 \node at (-100pt, 15pt){\textcolor{orange}{Heating}};
 \node at (-85pt, -99pt){\textcolor{orange}{Near phase transition}};
 \node at (-95pt, -215pt){\textcolor{orange}{Non-heating}};
 
    \end{tikzpicture}
    \caption{Comparison of entanglement entropy evolution for lattice-model and CFT calculations in the heating phase (top), near the phase transition (middle), and in the non-heating phase (bottom).  The lattice calculation is done in a system of length $L=600$ with periodical boundary conditions, and the subsystem is chosen as $A=[0,L/2]$. 
    From top to bottom, the driving parameters are the same as those in Fig.\ref{Fig:Lypidx_Epsl_Quasi}, respectively.
    }
 \label{Fig:EE_QuasiwithPT}
\end{figure}

These entanglement entropy evolutions can be verified by a free fermion lattice model calculation. See Sec.\ref{subSec:EE_evolution} for more details on the lattice models. Here we need to introduce the quasiperiodicity in $f(j)$ in \eqref{H1_lattice}. A sample plot of the lattice-model calculations of the entanglement entropy evolution is shown in Fig.\ref{Fig:EE_QuasiwithPT}, where one can find a good agreement with the CFT calculations.

\section{Discussion and conclusion}
\label{Sec:Discuss}

In this work, we have proposed a setup on quasi-periodically driven CFTs, where one can realize phase transitions between the heating and non-heating phases. The phase diagrams as well as the Lyapunov exponents can be analytically studied based on Avila's global theory of one-frequency quasiperiodic $\SL(2,\mathbb C)$ cocycles. In addition, based on Avila's theory, one can prove that there is no phase transition in the previously proposed setup on quasi-periodically driven CFTs \cite{wen2020periodically}. We verify our field theory results based on lattice model calculations and find a good agreement. 

\bigskip 

 Till now, we believe we have a good understanding on the phase diagrams of time-dependent driven CFTs with $sl(2,\mathbb R)$ deformations, as summarized in Table \ref{Phase_table}. However, the fate of time-dependent driven CFTs with general deformations, where the underlying algebra is the infinite-dimensional Virasoro algebra, is far from known. Recently, the \textit{periodically} driven CFTs with general deformations were studied in \cite{lapierre2020geometric,fan2020General}, where one observed both heating and non-heating phases based on specific choices of deformations. A systematic understanding of the phase diagram of the periodically driven CFTs, as well as its generalization to quasi-periodically/randomly driven CFTs is still needed.

As was explicitly shown in this work, the complexified Lyapunov exponents $\lambda_L(\epsilon)$ as well as the corresponding acceleration $\omega_\lambda$ play an important role in determining the phase diagram of the quasi-periodically driven CFTs. One natural question is: Is there any physical meaning of $\lambda_L(\epsilon)$ and $\omega_\lambda(\epsilon)$ in the driven CFT? 
It has been known that in the context of non-Hermitian quasi-crystals in the ground state, the quantized acceleration $\omega_\lambda$ corresponds to the winding number, which is a topological invariant characterizing the ground state of the non-Hermitian system \cite{2021_ChenShu,2023_Zhou_winding}.
For a system under time-dependent drivings, to our knowledge, the physical meaning of $\lambda_L(\epsilon)$ and $\omega_\lambda$ is not clear. In a forthcoming work \cite{Quantize_2025}, we will show that these quantities can be detected in driven CFTs as well as in the lattice model realizations. The price to pay is that we need to generalize the unitary time evolutions to \textit{non-unitary} time evolutions.

\acknowledgments

XW thanks Ruihua Fan, Yingfei Gu, and Ashvin Vishwanath for a closely related collaboration in \cite{wen2020periodically} and many helpful discussions.
This work is supported by a startup at Georgia Institute of Technology (JF, XW). QZ  was supported by National Key R\&D Program of China (2020 YFA0713300) and Nankai Zhide Foundation.

\appendix

\section{Building blocks of quasi-periodic driving: A single quantum quench}
\label{Appendix:BuildingBlock}

The quasiperiodic driving as considered in the main text is composed of a sequence of time steps (See, e.g., Fig.\ref{Fig:Driving_Protocol}). Within each time step, the driving Hamiltonian is fixed, and the problem is reduced to a single quantum quench.
In this appendix, we give details on the single quantum quench from the aspects of both CFT and lattice calculations.

One starts from the ground state of a homogeneous system with $f(x,t=0)=1$ in \eqref{H_deform_intro}. Then at $t=0$, one evolves the initial state with a deformed Hamiltonian $H_{\text{deform}}$.
To characterize the time evolution of wavefunction, we study the entanglement entropy in a subsystem $A$, say, defined in region $[kl, (k+1)l]$ where $k\in \mathbb Z$. That is, we choose the length of the subsystem as one wavelength of the deformation in \eqref{fx_sl2}. In this case, the entanglement entropy has a very simple expression \cite{wen2020periodically}
\be
S_A(t)-S_A(t=0)=\frac{c}{3}\log\big|\alpha(t)+\beta(t)\big|
+\frac{c}{3}\log\big|\alpha'(t)+\beta'(t)\big|.
\ee
The first (second) term on the right side of the above equation is contributed by the chiral (anti-chiral) component. 
For the Hamiltonian deformation considered in \eqref{fx_sl2}, the explicit forms of $\alpha(t)$ and $\beta(t)$ are given in \eqref{Elliptic}, \eqref{Parabolic}, and \eqref{Hyperbolic}, depending on the types of deformations.
For the expressions of $\alpha'(t)$ and $\beta'(t)$, which correspond to the contribution of the anti-chiral component, they are the same as $\alpha(t)$ and $\beta(t)$ except that one should change $a^- \to -a^-$.

\begin{figure}
\centering
\includegraphics[width=2.5in]{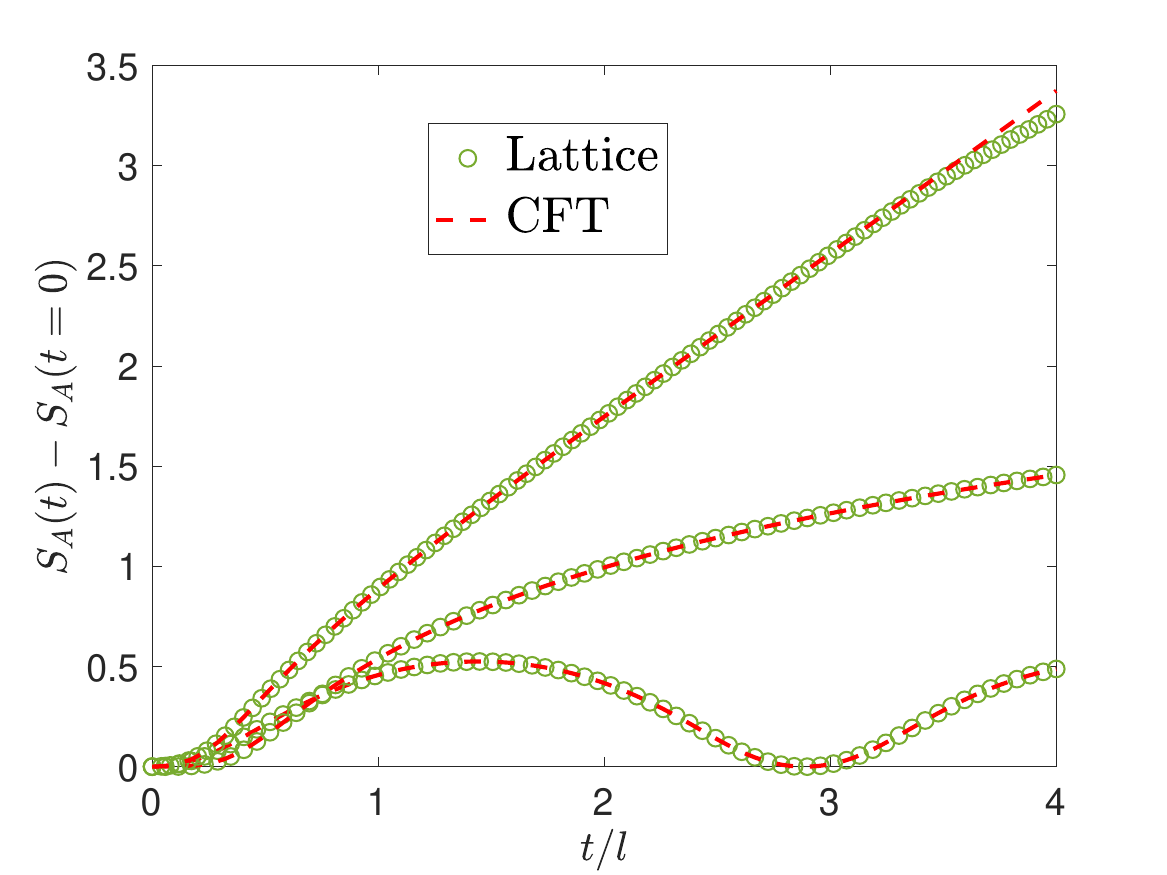}
	\caption{Entanglement entropy evolution for subsystem $A=[0,l]$ after three different types of quantum quenches by starting from the ground state of an un-deformed CFT Hamiltonian. From top to bottom, the quenched Hamiltonians are chosen as (i) hyperbolic type $(a^0,\,a^+,\,a^-)=(0.5,\, 0.2,\, 0.6)$, (ii) parabolic type $(a^0,\,a^+,\,a^-)=(0.5,\, 0,\, 0.5)$, and (iii) elliptic type $(a^0,\,a^+,\,a^-)=(0.5,\, 0.2,\, 0.3)$. In the lattice calculation, we choose $L=2l=600$.
	}
 \label{Fig:Quench}
\end{figure}

As shown in Fig.\ref{Fig:Quench}, we study the entanglement entropy evolution after a single quantum quench by considering all the three types of quenched Hamiltonians in \eqref{3Types}. For the lattice model calculations, see Sec.\ref{subSec:EE_evolution} for details.
One can find that the entanglement entropy will oscillate, grow logarithmically, and grow linearly in time if the quenched Hamiltonian is elliptic, parabolic, and hyperbolic, respectively.

\section{General cases on quasi-periodically driven CFT without phase transitions}
\label{Appendix:HeatingPhaseGerneral}

For the example considered in Sec.\ref{Sec:NoPhaseTransition}, we choose a uniform $H_0$, which is not deformed. In this appendix, we consider a general $H_0$ that is elliptic. 
As a remark, it is noted that for the type-I quasiperiodic driving, $H_0$ needs to be elliptic, since we do not know how to introduce the quasiperiodicity in time if $H_0$ is parabolic or hyperbolic, which is clear by looking at the expressions in \eqref{Elliptic}, \eqref{Parabolic}, and \eqref{Hyperbolic}. 

Let us first consider a simple deformation of $H_0$ that is easy to realize in lattice simulations \cite{wen2020periodically}, by choosing
\be
\label{tanh_deform_appendix}
a^0=1, \quad a^+=-\tanh(2\theta_0), \quad a^-=0,
\ee
The corresponding $\SU(1,1)$ matrix has the expression:
\be
\label{M0_theta_appendix}
\small
M_0(\theta_0,x)=\begin{pmatrix}
\cos x+i\cosh(2\theta_0)\sin x &-i \sinh(2\theta_0) \sin x\\
i\sinh(2\theta_0) \sin x &\cos x-i\cosh(2\theta_0) \sin x
\end{pmatrix},
\ee
where $x=\pi T_0/l_{\text{eff}}$, which is set to be quasiperiodic in time as $x=\pi \omega\cdot n$. Here we have chosen the driving time $T_0$ in the $n$-th driving cycle as $T_0=n\omega \, l_{\text{eff}}$ instead of $n\omega l$ as considered in the main text.
Then, by introducing 
\be
U_{\theta_0}=\begin{pmatrix}
\cosh \theta_0 &\sinh \theta_0\\
\sinh\theta_0 &\cosh\theta_0
\end{pmatrix},
\ee
one can check that $M_0(\theta_0,x)$ in \eqref{tanh_deform_appendix} and $M_0(\theta_0=0,x)$ in \eqref{Theta_0} are related by
\be
M_0(\theta_0,x)=U_{\theta_0} M_0(\theta=0,x) U_{\theta_0}^{-1}.
\ee
Then we have
\be
A(x):=M_0(\theta_0,x)M_1
=U_{\theta_0} \left[ M_0(0,x)\left( U_{\theta_0}^{-1}
M_1 U_{\theta_0}
\right)
\right] U_{\theta_0}^{-1}.
\nonumber
\ee
Since a constant conjugation does not change the Lyapunov exponent, then one can reduce it to the former case with $\theta_0=0$ in $M_0(\theta_0,x)$, and obtain
\be
\lambda_L(\epsilon=0)=\log|\tilde{\alpha}_1|,
\ee
where $\tilde\alpha_1$ is the first matrix elements in $ U_{\theta_0}^{-1}
M_1 U_{\theta_0}$, with the expression
\be
\label{tilde_alpha}
\small
\tilde\alpha_1=\Re (\alpha_1)+i\big[
\cosh(2\theta_0)\cdot \Im(\alpha_1)+\sinh(2\theta_0)\cdot \Im(\beta_1)
\big],
\ee
where $\alpha_1$ and $\beta_1$ are the elements in $M_1$ in \eqref{M1_general}. For example, if we consider the choice of $M_1$ in \eqref{M1_alpha_beta}, one can find 
\be
|\tilde\alpha_1|=\Big[\cos^2\left(\frac{\pi T_{1}}{l_{\text{eff}} }\right)+\sin^2\left(\frac{\pi T_{1}}{l_{\text{eff}}}\right)
\cdot 
\cosh^2(2\theta_0-2\theta)
\Big]^{1/2}.
\nonumber
\ee
We always have $|\tilde{\alpha}_1|>1$ as long as $M_1$ and $M_0$ do not commute with each other.

For general choices of $\alpha_1$ and $\beta_1$ in \eqref{tilde_alpha}, one can show that $|\tilde \alpha_1|>1$ as long as $M_0$ and $M_1$ do not commute. More explicitly, one can find that $|\tilde\alpha_1|^2\ge 1+[\cosh(2\theta_0)\cdot |\beta_1|-\sinh(2\theta_0)\Im(\alpha_1)]^2\ge 1$, where $|\tilde\alpha_1|^2=1$ if and only if $M_1$ commutes with $M_0$.
It is straightforward to check that the same conclusion also holds for a general elliptic $M_0(x)$.

Similar to the concrete examples studied in Sec.\ref{Sec:NoPhaseTransition}, if we use $H_0$ and $H_1$ above in the periodic driving \cite{han2020classification,wen2020periodically}, one can realize both heating and non-heating phases. However, by considering the same driving Hamiltonians and introducing quasiperiodicity in $T_0$, only the heating phase exists.

\section{General cases on quasi-periodically driven CFT with phase transitions}
\label{Appendix:PhaseTransition}

In Sec.\ref{Sec:PhaseTransition} of the main text, we illustrate the phase transitions in type-II quasiperiodic drivings
by choosing the quasiperiodic Hamiltonians to be elliptic. We have checked all the $3\times 3=9$ combinations of $H_0$ and $H_1$ based on the three types of Hamiltonians in \eqref{3Types}, and found that the phase transitions exist in each case. In this appendix, we give general results for all the other types of combinations. In all these cases,
the driving Hamiltonians $H_0$, as shown in Fig.\ref{Fig:Driving_Protocol} (bottom), depend on time in a quasiperiodic way.

\subsection{Quasiperiodic Hamiltonians that are elliptic}
\label{Appendix:case1}

\begin{figure}[tp]
\centering
\begin{tikzpicture}
  \node[inner sep=0pt] (russell) at (-65pt,-20pt)
    {\includegraphics[width=3.0in]{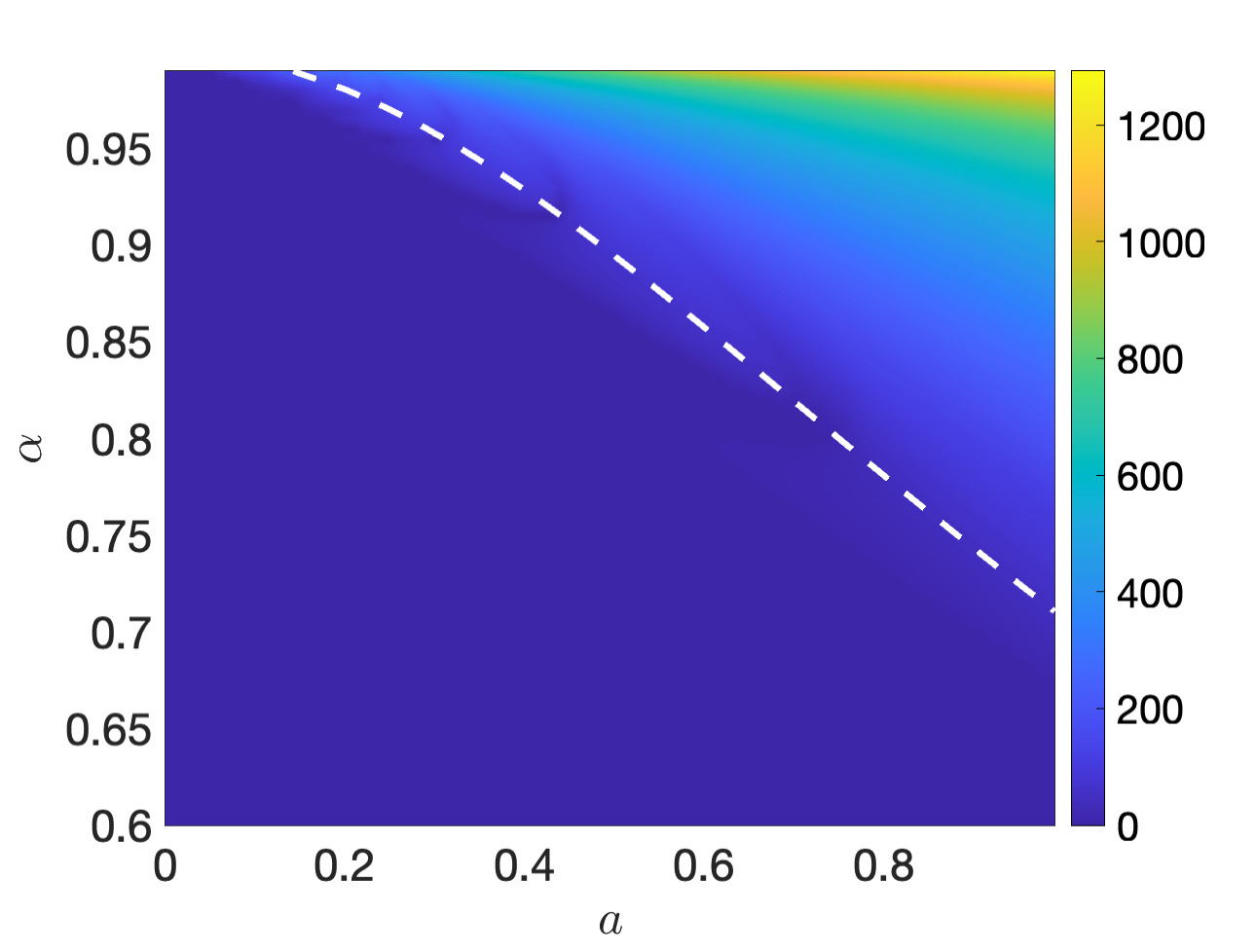}};
  \normalsize
 \node at (-100pt, -30pt){\textcolor{white}{Non-heating}};
 \node at (-35pt, +20pt){\textcolor{white}{Heating}};                   
    \end{tikzpicture}
    \caption{
Phase diagram of a quasi-periodically driven CFT as a function of the driving parameters $\alpha$ and $a$ in \eqref{Appendix_C_H0} and \eqref{M1_7}, respectively. The quasiperiodic frequency in \eqref{M0_7_appendix} is set to be $\omega=\pi(\sqrt{5}-1)/2$.
The white dashed line corresponds to the analytical result of phase transitions in \eqref{appendix_phaseTransition}.
The numerical values correspond to the entanglement entropy of  $A=[0,\,l]$ after $N=1000$ cycles of driving. 
}
 \label{Fig:Phase_appendix}
\end{figure}

\begin{figure}[htp]
\centering
\begin{tikzpicture}
  \node[inner sep=0pt] (russell) at (-65pt,-20pt)
    {\includegraphics[width=2.5in]{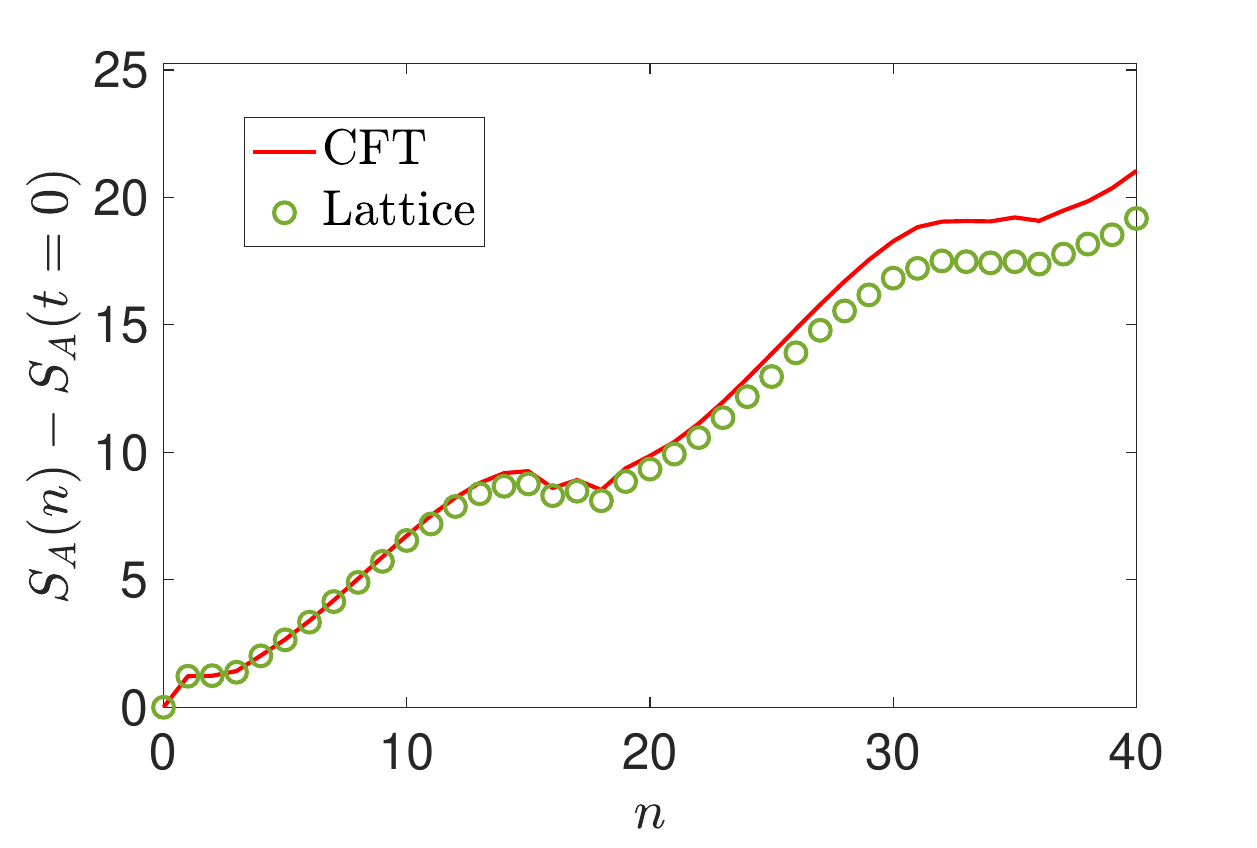}};

      \node[inner sep=0pt] (russell) at (-65pt,-140pt)
    {\includegraphics[width=2.5in]{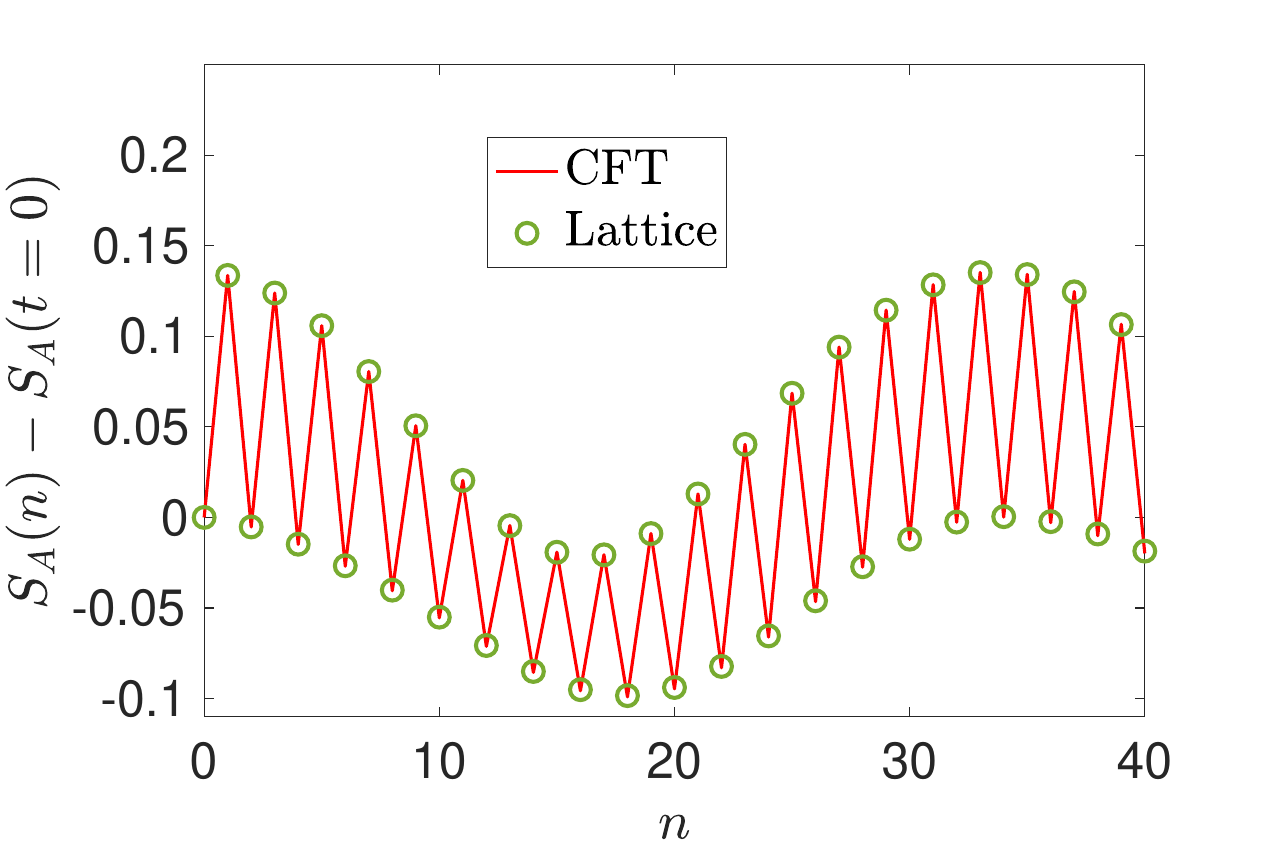}};
  
    \end{tikzpicture}
    \caption{Comparison of entanglement entropy evolution for lattice-model and CFT calculations in the heating phase (top) and in the non-heating phase (bottom), respectively.  The lattice calculation is done in a system of length $L=600$ with periodical boundary conditions, and the subsystem is chosen as $A=[0,L/2]$. 
    The driving parameters are chosen as $(a,\alpha)=(0.7, 0.95)$ in the heating phase (top), and $(0.2,\, 0.2)$ (bottom) which is deep in the non-heating phase in Fig.\ref{Fig:Phase_appendix}. 
    }
 \label{Fig:EE_QuasiwithPT_appendix}
\end{figure}

The driving Hamiltonians $H_0$ are elliptic, which are deformed quasi-periodically in time as
\be
\label{Appendix_C_H0}
a^0=1,\quad  a^+=\alpha \cos(\pi \omega\cdot n),\quad  a^-=\alpha\sin(\pi \omega\cdot n),
\ee
where $\alpha\in (0,\,1)$, $\omega$ is an irrational number, and $n\in \mathbb Z$. Here $n$ denotes the $n$-th time driving cycle.
We fix the driving time for each time step to be a constant $T_0=l_{\text{eff}}/2$, where $l_{\text{eff}}=l/\mathcal C_0$ is defined in \eqref{l_eff} and here we have $\mathcal C_0=\sqrt{1-\alpha^2}$. Then based on \eqref{Elliptic}, one can find the $\SU(1,1)$ matrix associated to $H_0$ in the $n$-th driving cycle is:
\be
\label{M0_7_appendix}
M_0=
\frac{i}{\mathcal C_0}\begin{pmatrix}
1 &\alpha \,e^{i\pi \omega\cdot n}\\
-\alpha\, e^{-i\pi \omega \cdot n} &-1
\end{pmatrix}.
\ee
Next, for $H_1$, we consider the very general form of $sl(2,\mathbb R)$ deformations.
The corresponding $\SU(1,1)$ matrix is of the general form in \eqref{M1_general}.
By denoting $x:=\pi \omega\cdot n$, we consider the corresponding cocycles $(\omega,\,A)$, where
\be
A(x):=M_0(x)M_1=\frac{i}{\mathcal C_0}\begin{pmatrix}
1 &\alpha \,e^{ix}\\
-\alpha\, e^{-ix} &-1
\end{pmatrix}
\begin{pmatrix}
\alpha_1 &\beta_1\\
\beta_1^\ast &\alpha_1^\ast
\end{pmatrix}.
\nonumber
\ee
Let us then complexify the phase by replacing $x$ with $x+i\epsilon$, and then let $\epsilon$ goes to positive infinity. One can obtain:
\be
\begin{split}
A(x+i\epsilon)=&\frac{i}{\mathcal C_0}e^\epsilon e^{-ix} 
\begin{pmatrix}
0 &0\\
-\alpha &0
\end{pmatrix}
\begin{pmatrix}
\alpha_1 &\beta_1\\
\beta_1^\ast &\alpha_1^\ast
\end{pmatrix}+\mathcal O(1)\\
=&
\frac{i}{\mathcal C_0}e^\epsilon e^{-ix} 
\begin{pmatrix}
0 &0\\
-\alpha\alpha_1 &-\alpha\beta_1
\end{pmatrix}
+\mathcal O(1).
\end{split}
\ee
The constant matrix $B=\begin{pmatrix}
0 &0\\
-\alpha\alpha_1 &-\alpha\beta_1
\end{pmatrix}$ has Lyapunov exponent:
\be
\lim_{n\to \infty}
\frac{\log||B^n ||}{n}=
\lambda_{\text{max}}(B)=\log\left|
\frac{\alpha \beta_1}{\sqrt{1-\alpha^2}}
\right|.
\ee
Therefore, we have $\lambda_L(\epsilon)=\epsilon+\log\left|
\frac{\alpha \beta_1}{\sqrt{1-\alpha^2}}
\right|+\mathcal O(1)$. Avila's global theory shows that as a function of $\epsilon$, $\lambda_L(\epsilon)$ is a convex, piecewise linear function, and their slopes are integers. This implies that
\be
\lambda_L(\epsilon)=\max\left\{
\epsilon+\log\left|
\frac{\alpha \beta_1}{\sqrt{1-\alpha^2}}
\right|,\,\lambda_0(A)
\right\}.
\ee
Similarly as what we describe in the main text, the cocycle $(\omega,A)$ is uniformly hyperbolic if and only if $\lambda_0(A)>0$ and $\omega_\lambda(\epsilon)=0$ for $\epsilon=0^+$. If $(\omega,A)$ is not uniformly hyperbolic, which means $(\omega,A)$ is in one of the three cases in Fig.\ref{Lyapunov_Phase}, then we have 
\be
\lambda_L(\epsilon=0)=\max\left\{
\log\left|
\frac{\alpha \beta_1}{\sqrt{1-\alpha^2}}
\right|,\, 0
\right\}.
\ee

To give an illustration, we consider the specific choice of driving Hamiltonian $H_1$, which is hyperbolic, by using the following deformation:
\be
\label{appendix_H1_hyperbolic}
a^0=a^+=0, \quad a^-=-1.
\ee
With this deformation, one has $\mathcal C=1$ and  $l_{\text{eff}}=l/\mathcal C=l$. Based on \eqref{Hyperbolic}, one can find the corresponding $\SU(1,1)$ matrix as
\be
\label{M1_7}
M_1=\begin{pmatrix}
\cosh\left(\frac{\pi t}{l}\right) &\sinh\left(\frac{\pi t}{l}\right) \\
\sinh\left(\frac{\pi t}{l}\right)  &\cosh\left(\frac{\pi t}{l}\right) 
\end{pmatrix}
=:
\begin{pmatrix}
\sqrt{1+a^2} &a \\
a &\sqrt{1+a^2}
\end{pmatrix},
\ee
where we have defined $a=\sinh(\pi t/l)$. Then based on the above discussion, if $(\omega,A)$ is not uniformly hyperbolic, we will have
\be
\lambda_L(\epsilon=0)=\max\left\{
\log\frac{\alpha\cdot a}{\sqrt{1-\alpha^2}},\,0
\right\}
\ee
where the heating phase is determined by 
\be
 \log\frac{\alpha\cdot a}{\sqrt{1-\alpha^2}}>0, \quad \text{ i.e., }\quad 
 \alpha^2(1+a^2)>1,
\ee
and the phase transitions to the non-heating phases are determined by 
\be
\label{appendix_phaseTransition}
\alpha^2(1+a^2)=1,
\ee
which corresponds to the white dashed line in the two dimensional parameter space spanned by $a$ and $\alpha$
in Fig.\ref{Fig:Phase_appendix}.
By comparing with the CFT calculation of the entanglement entropy growth in Fig.\ref{Fig:Phase_appendix}, one can find that the analytical result in \eqref{appendix_phaseTransition} gives a good description of the phase transitions.
Once we fix the phase transitions, we can study the entanglement entropy evolution in the heating or non-heating phases, as shown in Fig.\ref{Fig:EE_QuasiwithPT_appendix}. 
One can see a linear growth of EE in the heating phase (the feature of linear growth will become clearer if we take a longer time of driving), and an oscillation of EE with no growth in the non-heating phase.

\subsection{Quasiperiodic Hamiltonians that are parabolic}

Now we consider the parabolic Hamiltonians $H_0$, which are deformed quasi-periodically in time as
\be
\label{appendix_H0_parabolic}
a^0=1, \quad a^+=\cos(\pi \omega\cdot n),\quad a^-=\sin(\pi \omega\cdot n),
\ee
where $n$ denotes the $n$-th driving cycle.
The corresponding $\SU(1,1)$ matrix in the $n$-th driving cycle has the expression
\be
M_0(n)=
\begin{pmatrix}
1+i\alpha &i\alpha\, e^{i\pi \omega\cdot n}\\
-i\alpha\, e^{-i\pi \omega\cdot n} &1-i\alpha
\end{pmatrix},
\ee
where we have defined $\alpha=\pi T_0/l$.
For $H_1$, we consider the very general form of 
$sl(2,\mathbb R)$ deformations, and 
the corresponding $\SU(1,1)$ matrix is of the general form in \eqref{M1_general}.

Then we can consider the cocycle $(\omega, A)$, with
\be
\label{appendix_Ax_parabolic}
A(x):=M_0(x)M_1=\begin{pmatrix}
1+i\alpha &i\alpha\, e^{ix}\\
-i\alpha\, e^{-ix} &1-i\alpha
\end{pmatrix}
\begin{pmatrix}
\alpha_1 &\beta_1\\
\beta_1^\ast &\alpha_1^\ast
\end{pmatrix},
\ee
where we have defined $x=\pi\omega\cdot n$.

By performing a similar calculation and the same reasoning as in the previous subsection, one can find that if $(\omega, A)$ is not uniformly hyperbolic, then we have
\be
\lambda_L(\epsilon=0)=\max\left\{
\log|\alpha \beta_1|,\,0
\right\}.
\ee

Including the case where $H_0$ is hyperbolic, which is studied in the main text, we have tested all the  $3\times 3=9$ combinations of $H_0$ and $H_1$ in the type-II quasiperiodic driving. The phase transitions can be observed in all these cases, which indicates that the phase transition is a generic feature in our type-II quasiperiodic drivings.

\section{A remark on the difference between the uniformly and non-uniformly hyperbolic cases}
\label{Uniform_non-uniform}

\begin{figure}[tp]
\centering
\begin{tikzpicture}
  \node[inner sep=0pt] (russell) at (-65pt,-20pt)
    {\includegraphics[width=2.8in]{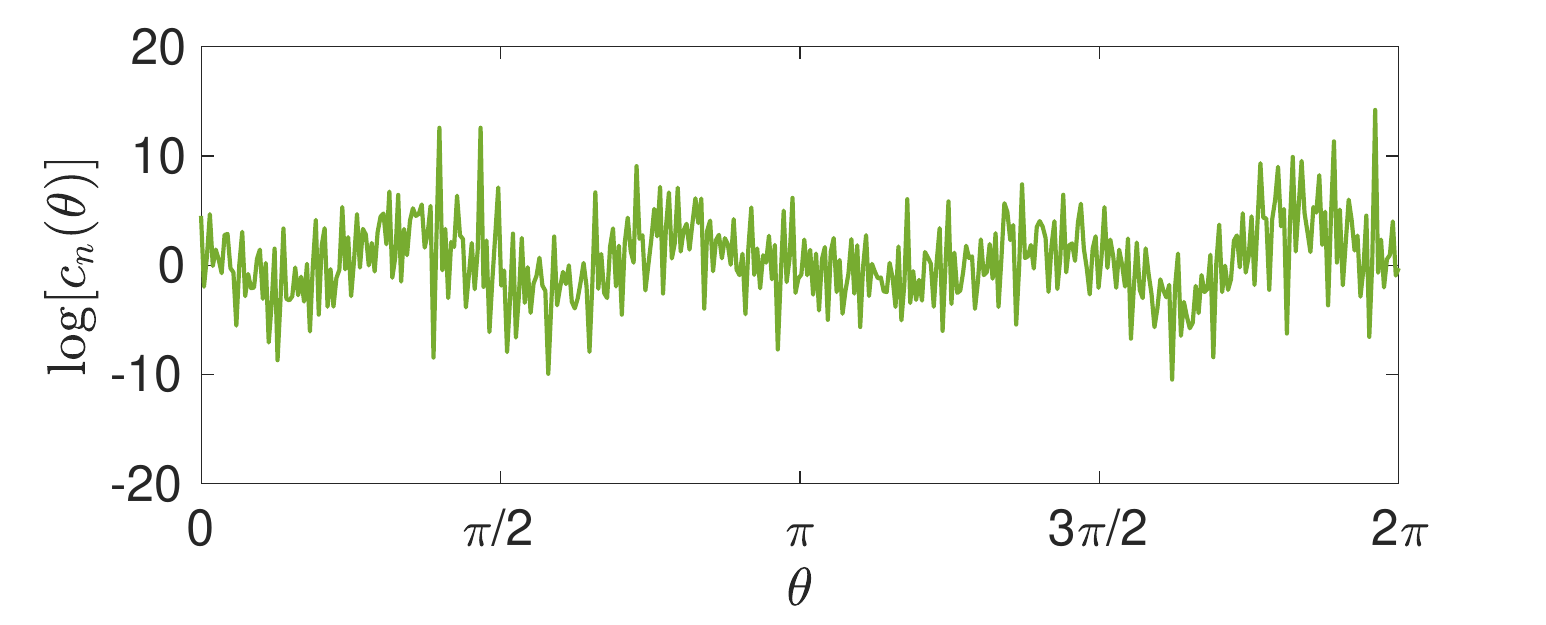}};

      \node[inner sep=0pt] (russell) at (-65pt,-105pt)
    {\includegraphics[width=2.8in]{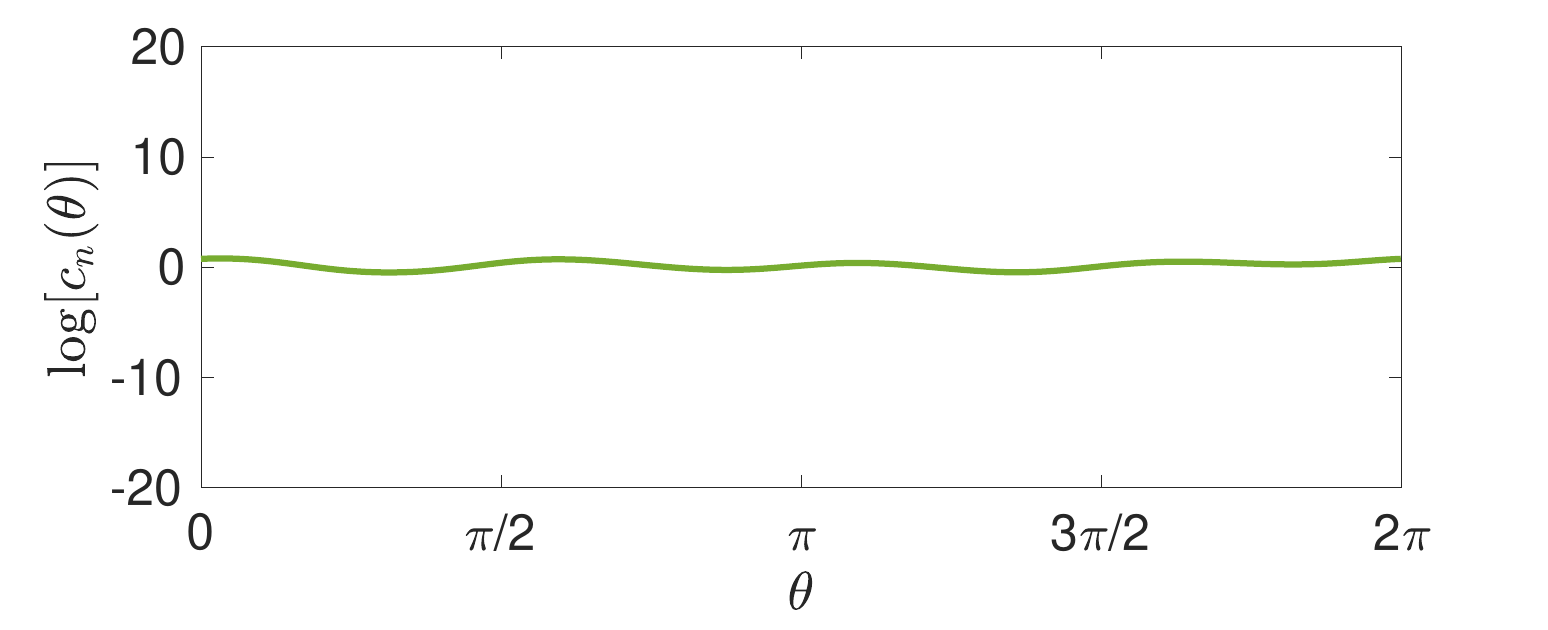}};
 \node at (-70pt, 8pt){\textcolor{orange}{Non-uniformly hyperbolic}};
  \node at (-80pt, -78pt){\textcolor{orange}{Uniformly hyperbolic}};
                  
    \end{tikzpicture}
    \caption{
    Comparison of $\log[c_n(\theta)]$ for non-uniformly hyperbolic (top) and uniformly hyperbolic (bottom) cases based on a numerical calculation evaluation of \eqref{eq:An_NUH_UH}.
    The parameters we use are $n=5\times 10^5$, with deformations $\alpha=1.2, a=2$ (top), and $\alpha=0.9, a=1.5$ (bottom). These two cases correspond to non-uniformly and uniformly hyperbolic cocycles based on their features in the complexified Lyapunov exponents $\lambda_L(\epsilon)$ and the acceleration $\omega_\lambda$.
    $c_n(\theta)$ has a strong fluctuation and depends on $\theta$ in a ``non-uniform'' way in the non-uniformly hyperbolic case.
    }
 \label{Fig:uniform_non-uniform}
\end{figure}

As discussed in Sec.\ref{Subsec:Avila} and Sec.\ref{Sec:PhaseTransition} in the main text,
if $\lambda_L>0$, the cocycles could be either uniform or nonuniform hyperbolicity.
One can distinguish these two cases by considering the complexified Lyapunov exponents and the corresponding accelerations, as shown in Fig.\ref{Lyapunov_Phase} and Fig.\ref{Fig:Uniform_Hyperbolic}. In this appendix, we introduce another way to distinguish these two cases without complexifying the cocycles. 

By definition, the cocycle $(\omega,A)$ as introduced in Sec.\ref{Subsec:Avila} is \textit{uniformly hyperbolic} if there exists a continuous splitting $E_s(x)\oplus E_u(x)=\mathbb R^2$, and $C>0$, $0<\lambda<1$ such that for every $n\ge 1$ and $x\in \Sigma$ (where $\Sigma=S^1$ in our setup) we have \cite{viana2014lectures}
\be
\begin{split}
||A_n(x)\cdot w ||\le C\lambda^n || w||, \quad w\in E_s(x),\\
||A_{-n}(x)\cdot w ||\le C\lambda^n || w||, \quad w\in E_u(x).
\end{split}
\ee
Clearly every uniformly hyperbolic cocycle has positive Lyapunov exponent.

For $\SL(2,\mathbb R)$ or $\SU(1,1)$ cocycles, a more handy criterion to determine whether a cocycle is uniformly hyperbolic is available: For uniformly hyperbolic cocycles, it is enough to find constants $C,\,\lambda>0$ such that for every $x\in \Sigma$
(where $\Sigma=S^1$ in our setup) and $n\ge 1$, we have \cite{yoccoz}
\be
\label{UniformH_criterion}
||A_n(x) ||\ge C \,e^{\lambda n}.
\ee

Based on the above criteria, we can distinguish the uniformly and non-uniformly hyperbolic cocycles by studying the exponential growth $||A_n(x)||\simeq c(x)\, e^{\lambda n}$ as we gradually change $x$ over $\Sigma=S^1$.
If the cocycle is non-uniformly hyperbolic, the coefficient $c(x)$ will fluctuate strongly. Indeed, there is a famous result of Man\'e \cite{mane1977quasi,viana2014lectures}, that if a cocycle is not
uniformly hyperbolic then there always exists a vector that is never expanded, either in the future or in the
past.
On the other hand, for a uniformly hyperbolic cocycle, it is expected that the coefficient $c(x)$ changes smoothly with $x\in S^1$ and has a lower bound.

Now, let us give a concrete example to illustrate this difference.
We consider the type-II quasiperiodic driving where the family of $H_0$ are parabolic and $H_1$ is hyperbolic.
More explicitly, $H_0$ are obtained from the deformations in \eqref{appendix_H0_parabolic}, and $H_1$ is obtained from the deformation in \eqref{appendix_H1_hyperbolic}. Then we can define $A(x)$ according to \eqref{appendix_Ax_parabolic}, and generalize it by considering $x\to x+\theta$, where $\theta\in[0,2\pi)$. If the cocycles are hyperbolic (no matter uniformly or non-uniformly), we will have
\be
\label{eq:An_NUH_UH}
|| A_n(x+\theta)||\simeq c_n(\theta)\, e^{\lambda_L \cdot n}, \quad n\to \infty,
\ee
where $\theta\in[0,2\pi)$.
Here the coefficient $c_n(\theta)$ will exhibit different features for non-uniformly hyperbolic and uniformly hyperbolic cases. As shown in Fig.\ref{Fig:uniform_non-uniform}, we plot $\log[c_n(\theta)]$ as a function of $\theta$ for both non-uniformly and uniformly hyperbolic cases. One can see clearly that $\log[c_n(\theta)]$ have a strong fluctuation and depend on $\theta$ in a ``non-uniform'' way in the non-uniformly hyperbolic case. For uniformly hyperbolic case, $\log[c_n(\theta)]$ depend on $\theta$ smoothly in a ``uniform'' way.

In terms of entanglement entropy evolution, while the leading term always grows as $\lambda_L\cdot n$ for both non-uniformly and uniformly hyperbolic cases, the different features of $\log[c_n(\theta)]$ in Fig.\ref{Fig:uniform_non-uniform} will affect the subleading term of the entanglement entropy. That is, for the non-uniformly hyperbolic case, the subleading term of entanglement entropy evolution will have a much stronger fluctuation than that in the uniformly hyperbolic case.

\bibliography{QuasiCFT}

\end{document}